\documentclass{article}
\usepackage{epsfig} 
\usepackage{eufrak}
\usepackage{overcite}
\DeclareSymbolFont{rsfscript}{OMS}{rsfs}{m}{n}
\DeclareSymbolFontAlphabet{\mathrsfs}{rsfscript}

\newlength{\normalarraycolsep}
\newlength{\normaltabcolsep}
\newcommand{\dd}{\mathrm{d}}
\newcommand{\lb}{\ell_B}

\newcommand{\nb}{n_\mathrm{b}}
\newcommand{\nc}{n_\mathrm{c}}

\newcommand{\kb}{\kappa_\mathrm{b}}

\newcommand{\arcsinh}{\mathop\mathrm{arcsinh}\nolimits}
\newcommand{\arccosh}{\mathop\mathrm{arccosh}\nolimits}
\newcommand{\arctanh}{\mathop\mathrm{arctanh}\nolimits}
\newcommand{\arccotg}{\mathop\mathrm{arccotg}\nolimits}
\newcommand{\sn}{\mathop\mathrm{sn}\nolimits}
\newcommand{\cn}{\mathop\mathrm{cn}\nolimits}
\newcommand{\cs}{\mathop\mathrm{cs}\nolimits}

\newcommand{\be}{\begin{equation}}
\newcommand{\ee}{\ \ \ \end{equation}}
\newcommand{\bea}{\setlength{\arraycolsep}{0.4\normalarraycolsep}
                  \begin{eqnarray}} 
\newcommand{\eea}{\ \ \ \end{eqnarray}\setlength{\arraycolsep}
                  {\normalarraycolsep}}
\newcommand{\bean}{\setlength{\arraycolsep}{0.4\normalarraycolsep} 
                  \begin{eqnarray*}}
\newcommand{\eean}{\ \ \ \end{eqnarray*}\setlength{\arraycolsep}
                  {\normalarraycolsep}}

\begin{document}
\title{Where the linearized Poisson-Boltzmann cell model fails: (II) the
planar case as a prototype study}
\author{
M. N. Tamashiro and H. Schiessel\\ 
\textit{Max-Planck-Institut f\"ur Polymerforschung,}\\
\textit{Ackermannweg 10, 55128 Mainz, Germany}
}
\date{}
\maketitle
\begin{abstract}
The classical problem of two uniformly charged infinite  planes 
in electrochemical equilibrium with an infinite monovalent 
salt reservoir is solved exactly at the mean-field
nonlinear Poisson-Boltzmann (PB) level, including an 
explicit expression of the associated nonlinear
 electrostatic contribution to the 
 semi-grand-canonical potential.
A linearization of the nonlinear functional is presented 
that leads to Debye-H\"uckel-like equations agreeing asymptotically with
the nonlinear PB results in the 
weak-coupling (high-temperature) and 
counterionic ideal-gas limits. 
This linearization scheme yields
artifacts in the low-temperature, large-separation
or high-surface charge limits.
In particular, the osmotic-pressure difference between 
the interplane region and the salt reservoir
becomes negative in the above limits, in disagreement with the 
exact (at mean-field level) nonlinear PB solution. 
By using explicitly gauge-invariant forms of the electrostatic potential 
we show that these artifacts --- 
although thermodynamically consistent with quadratic 
expansions of the nonlinear functional ---  
can be traced back to the non-fulfillment of the 
underlying assumptions of the linearization.
Explicit comparison between the analytical expressions of the
exact nonlinear solution and the corresponding linearized 
equations allows us to show that the linearized results are 
asymptotically exact in the weak-coupling and 
counterionic ideal-gas limits, 
but always fail otherwise, predicting negative osmotic-pressure differences. 
By taking appropriate limits of
the full nonlinear PB solution, we provide 
asymptotic expressions for the semi-grand-canonical potential and 
the osmotic-pressure difference that involve 
only elementary functions, which cover the complementary 
region where the linearized theory breaks down.
\end{abstract}

\section{Introduction}

In the preceding paper \cite{sphere}
we performed a linearization of the
 mean-field Poisson-Boltzmann (PB) density functional
for spherical Wigner-Seitz cells that agrees asymptotically 
with the PB results in the weak-coupling (high-temperature) 
limit by adopting an explicitly gauge-invariant approach. 
However, as already pointed out previously 
in the literature \cite{grunberg,deserno},
the linearization scheme yields artifacts 
in the low-temperature, high-surface charge or infinite-dilution
(of polyions) limits for the Donnan 
equilibrium problem \cite{donnan,overbeekdonnan,hill1,reus,tamashiro} 
 in spherical geometry, which describes a suspension of spherical 
charged polyions in electrochemical 
equilibrium with an infinite salt reservoir. In these limits the linearized 
osmotic-pressure difference between the colloidal suspension and the 
salt reservoir becomes negative, in disagreement with the full 
nonlinear PB result, that always displays positive 
osmotic-pressure differences.
Because the nonlinear PB equation is 
not analytically solvable in spherical geometry 
  --- even in the simplest salt-free case,
when only neutralizing counterions are present --- 
one must rely on numerical 
calculations to establish comparisons between the nonlinear and
the linearized equations. This motivated us to consider the 
prototype case represented by the classical problem of two uniformly 
charged infinite planes in electrochemical equilibrium with an infinite salt 
reservoir, when the exact analytical solution of the nonlinear problem is
possible. The explicit analytical comparison between the exact (at
the mean-field level) full nonlinear and the approximated linearized 
equations allows us to trace back the underlying 
reasons of the breakdown of
the linearization scheme that is intrinsically 
associated with its range of validity.  
Additionally, the study of this exactly-solvable case 
sheds some light on the question of the proper definition of the 
linearized osmotic pressure that was previously considered 
in Ref.~[\citen{deserno}].

Moreover, to our knowledge, the explicit calculation 
of the semi-grand-canonical potential for two uniformly 
charged infinite planes 
at the nonlinear mean-field PB level has only been reported 
in connection to the polyelectrolyte-brush 
problem \cite{brushes}. 
In that work, however, the thermodynamical potential 
also included electrostatic and elastic contributions 
arising from the polyelectrolyte brushes, and therefore, 
these need to be subtracted out.
The knowledge of the thermodynamic potential allows us to derive 
all thermodynamic properties of the two charged infinite planes problem.
We note that it can be also extended to curved surfaces by using 
the Derjaguin 
approximation \cite{derjaguin,white,israelachvili}. It is then possible 
to determine the normal forces (per unit area) between these  
surfaces when their separation distance is much smaller 
than their curvature radius. We will present the exact nonlinear
semi-grand-canonical functional from which we derive 
approximate expressions. These involve only elementary functions and 
provide excellent approximations to the full nonlinear PB results 
within the whole range of parameters. 

The remainder of the paper is organized as follows. 
In Section~\ref{section:2} the model is introduced and 
the exact nonlinear solution is obtained. 
In Section~\ref{section:3} the linearization of the 
appropriate semi-grand-canonical functional is performed.
In Section~\ref{section:4} we present explicit analytic 
comparisons between the exact nonlinear and the 
linearized equations of state. 
In Section~\ref{section:5} we give some concluding
remarks. 
In Appendix~\ref{appendix:a} we compare the exact 
nonlinear and the self-consistent linearized 
averaged densities used to perform the 
quadratic expansion of the nonlinear functional. 
In Appendix~\ref{appendix:b} we derive 
the asymptotic expansions of the exact nonlinear 
solution. 
In Appendix~\ref{appendix:c} we obtain extended expansions
of the exact nonlinear solution that are valid in the 
Gouy-Chapman (high-surface charge) and the large-separation limits, 
regions where the linearization scheme breaks down. 
In Appendix~\ref{appendix:d} we present 
linearized equations that preserve the exactness of the 
Legendre transformation between the semi-grand-canonical and 
canonical-ensemble formulations of the problem. 

\section{Exact nonlinear Poisson-Boltzmann solution\label{section:2}}

The system to be considered is comprised of two infinite 
planar surfaces a distance $2L$ apart, each 
with surface charge $-\sigma q$, where $q>0$ is the 
elementary charge, in electrochemical equilibrium with an infinite
monovalent salt reservoir of bulk salt density $\nb$.  
The microions (positive counterions and salt ions) 
are free to move in the region $-L< x < L$
between the two charged surfaces, 
where we introduced a Cartesian axis perpendicular to the planes in
which the midplane is located at $x=0$ and the two charged 
planes at $x=\pm L$. 
At the mean-field level of approximation the ions are treated as 
inhomogeneous ideal gases described by their average local 
number densities $n_\pm(x)$.
We do not distinguish between (positive) counterions and positive 
ions derived from the salt dissociation. 
The total charge number density (counterions, salt ions and the negative 
surface charge on the planes) of the system, 
\be
\rho(x)=n_+(x)-n_-(x)-\sigma\delta(x+L)-\sigma\delta(x-L),
\ee
where $\delta$ is the one-dimensional Dirac delta-function,
is related  to the reduced electrostatic potential 
$\psi(x)\equiv\beta q \Psi(x)$, which satisfies 
the (exact) Poisson equation,
\be
\frac{\dd^2\psi(x)}{\dd x^2} = -4\pi\lb \rho(x), \label{eqn:poisson}
\ee
where $\lb\equiv\beta q^2/\epsilon$
is the Bjerrum length and $\beta^{-1}=k_B T$ is the thermal energy. 
It is implicitly assumed that the solvent dielectric constant 
$\epsilon$ remains the same outside the region containing the salt
solution $(|x|>L)$, so image-charge effects due 
to dielectric contrast are absent.
The formal solution of the Poisson equation~(\ref{eqn:poisson}) 
may be written in
terms of the one-dimensional Green function $G_1(x,x_0)$,
\be
\psi(x)=\lim_{\delta L\to 0_+}\lb \int
\limits_{L-\delta L}^{L+\delta L}\dd x_0\,G_1(x,x_0)\,\rho(x_0) ,
\qquad  \frac{\dd^2 G_1(x,x_0)}{\dd x^2} = -4\pi\delta(x-x_0), 
\ee
which in turn allows us to express the mean-field
semi-grand-canonical functional (for one charged plane) per unit area as
\bea
\frac{\beta\mathit{\Omega}}{A}&=&
\lim_{\delta L\to 0_+}\!\frac{\lb}2\!\!\int\limits_0^{L+\delta L}
\!\!\dd x
\int\limits_0^{L+\delta L} 
\!\!\dd x_0\,
\rho(x)\,G_1(x,x_0)\,\rho(x_0)
+ \sum_{i=\pm}\int\limits_0^L\dd x\,n_i(x)\left\{\ln
  \left[{n_i(x)}\zeta_i^3\right]-1-\beta\mu_i\right\}\nonumber\\
&=&\frac1{8\pi\lb}\int\limits_0^L\dd x\left[\frac{\dd\psi(x)}{\dd x}\right]^2
+ \sum_{i=\pm}\int\limits_0^L\dd x\,n_i(x)\left\{\ln
  \left[\frac{n_i(x)}{\nb}\right]-1\right\} 
, \quad\label{eqn:omegapb}
\eea
where $\zeta_\pm$ are the thermal de Broglie wavelengths of 
 cations (including the positive counterions)
and anions, respectively, 
and the (mean-field) microion 
chemical potentials 
$\beta\mu_\pm=\ln \left(\nb \zeta^3_\pm\right)$
assume ideal gases of uniform density $\nb$ for both types of ions in 
the infinite salt reservoir. The last equality results 
from integrating the electrostatic energy (per unit area) --- 
the first term of Eq.~(\ref{eqn:omegapb}) ---  by parts
and using the fact that surface contributions 
vanish due to Gauss' law and the overall electroneutrality
of the system,  
\be
\lim_{\delta L\to 0_+}\int\limits_0^{L+\delta L}\dd x\, \rho(x) =0 , 
\qquad\quad\mbox{ or }   
\qquad\quad
\int\limits_0^L\dd x\, \left[n_+(x)-n_-(x)\right] = \sigma .  
\label{eqn:neut_planes}
\ee

The equilibrium density profiles are obtained by minimizing 
the PB semi-grand-canonical functional~(\ref{eqn:omegapb})
under the charge-neutrality constraint~(\ref{eqn:neut_planes}),
\be
\frac{\delta}{\delta n_\pm(x)}
\left[\frac{{\mathit\Omega}}{A}-
\mu_\mathrm{el}\lim_{\delta L\to 0_+}\int\limits_0^{L+\delta L}
\dd x\,\rho(x)\right] = 0, 
\qquad
\ee
where we introduced the Lagrange multiplier $\mu_\mathrm{el}$
to ensure the neutrality condition~(\ref{eqn:neut_planes}). 
This yields the Boltzmann-weighted ionic profiles,
\be
n_\pm(x)=\nb\exp\left[\pm\beta\mu_\mathrm{el}\mp \psi(x)\right], 
\label{eqn:planeprofile}
\ee
and the Lagrange multiplier $\mu_\mathrm{el}$ is 
found by imposing the charge-neutrality 
condition~(\ref{eqn:neut_planes}),
\bea
\mathrm{e}^{\pm\beta\mu_\mathrm{el}} 
&=&\frac{\sqrt{\nc^2 +
(2\nb)^2\alpha_+\alpha_-}\pm \nc}{2\nb \alpha_\pm}\,
\mathrm{e}^{\pm\left\langle\psi\right\rangle } , 
\label{eqn:muel} \\
\alpha_\pm &=&\left\langle 
\mathrm{e}^{\pm \left\langle\psi\right\rangle \mp \psi(x)}\right\rangle, 
\eea
where $\nc=\sigma/L$ is the average density of counterions in the 
interplane region $|x|< L$ and  
the brackets denote unweighted spatial averages over the volume 
available to the microions, 
\be
\left\langle {\cal X}(x)\right\rangle\equiv 
\frac1L \int\limits_0^L \dd x\, {\cal X}(x) . 
\ee
In close analogy to the 
spherical case --- cf.~Appendix~E of Ref.~[\citen{sphere}] ---  
it is possible to write the nonlinear 
equilibrium density 
profiles in an explicitly  
gauge-invariant form by inserting the 
Lagrange multiplier~(\ref{eqn:muel}) 
into the Boltzmann-weighted ionic profiles,
Eqs.~(\ref{eqn:planeprofile}),
\bea
n_\pm(x)&=& \frac{\sqrt{\nc^2 +
(2\nb)^2\alpha_+\alpha_-}\pm \nc}{2\alpha_\pm}\, 
\mathrm{e}^{\pm\left\langle\psi\right\rangle \mp \psi(x)}.
\label{eqn:npmgauge}
\eea
By explicitly gauge-invariant we mean that the equilibrium profiles 
do not depend \textit{explicitly}
on a particular choice of the zero of the potential,
because they depend only on the difference 
$\left\langle\psi\right\rangle-\psi(x)$.
Henceforth \textit{gauge-invariant} will be a short writing 
to \textit{explicitly gauge-invariant}.
In particular, in the salt-free $(\nb\to 0)$ limit, these
gauge-invariant forms lead --- in a direct and transparent 
way --- to the vanishing coion profile, $n_-(x)\equiv 0$, 
and to the salt-free equilibrium counterion profile, 
$n_+(x)=\nc \mathrm{e}^{\left\langle\psi\right\rangle
  -\psi(x)}/\alpha_+$. 

The most commonly used gauge \cite{israelachvili,andelman} 
is the one in which the
charge-neutrality Lagrange multiplier is zero, 
$\mu_\mathrm{el}\equiv 0$, 
which \textit{does not correspond} to the gauge in which 
the electrostatic potential \textit{at the infinite salt reservoir} 
vanishes \cite{footnote1}.
Henceforth, to simplify the notation and the calculations 
we will use the standard gauge $\mu_\mathrm{el}\equiv 0$
to treat the \textit{nonlinear problem.} 
We should keep in mind, however, that 
the fixed-gauge electrostatic 
potential $\varphi(x)\equiv \psi(x)-\beta\mu_\mathrm{el}$ 
will no longer be gauge-invariant: its value at a particular 
point ---  let us say, at the midplane, 
 $\varphi_0\equiv\varphi(x=0)$, 
or at the charged surfaces, $\varphi_L\equiv\varphi(x=L)$ ---
 will be determined by imposing the overall charge 
neutrality~(\ref{eqn:neut_planes}) in the whole system. 
They can no longer be chosen arbitrarily, in contrast to
their gauge-invariant counterparts, $\psi_0\equiv\psi(x=0)$ or  
$\psi_L\equiv\psi(x=L)$. In the gauge-invariant formulation,  
either $\psi_0$ or $\psi_L$ may be 
chosen arbitrarily --- but not both \textit{simultaneously} --- 
because the difference 
$\psi_L-\psi_0= \varphi_L- \varphi_0$ must eventually be preserved. 

In the standard gauge the nonlinear problem 
reduces into solving the usual PB 
equation for two charged infinite 
planes\cite{israelachvili,andelman,footnote2}, 
\be
\frac{\dd^2 \varphi(x)}{\dd x^2} = \kb^2 \sinh\varphi(x)
+ \frac2{\Lambda}\left[\delta(x+L) + \delta(x-L)\right],
\qquad\quad n_\pm(x)= \nb \mathrm{e}^{\mp \varphi(x)},\quad
\label{eqn:nonlinearpb}
\ee
with the appropriate boundary conditions, 
\be
\psi'(x=0)=\varphi'(x=0)=0,\qquad\mbox{ and } \qquad
\psi'(x=\mp L)=\varphi'(x=\mp L)= \pm \frac2{\Lambda},
\label{eqn:boundarycond}
\ee
the prime $(')$ denoting differentiation
with respect to the argument.
We have introduced two length scales:  
the Debye screening length associated with the bulk 
density $\nb$ of the infinite salt reservoir, 
\be
\kb^{-1}\equiv \frac1{\sqrt{8\pi\lb\nb}}, 
\ee
and the Gouy-Chapman \cite{gouy,chapman} length,
\be
\Lambda\equiv\frac{1}{2\pi\lb \sigma},
\ee
which gives the characteristic (algebraic) 
decay length of the counterion distribution 
(for a salt-free system) 
around an infinite charged plane with bare surface charge $\sigma$.

Using the mathematical identity
\be
\frac{\dd^2 \varphi(x)}{\dd x^2} =\frac12\,\frac{\dd (\varphi')^2}{\dd \varphi},
\ee 
it is possible to integrate the nonlinear PB 
Eq.~(\ref{eqn:nonlinearpb}) exactly,
\bea
\left[\varphi'(x)\right]^2&=& \kb^2\left[2\cosh\varphi(x)-2\cosh\varphi_0\right], \\
\kb |x| &=& \int\limits_{\varphi(x)}^{\varphi_0} \frac{\dd\varphi}
{\sqrt{2\cosh\varphi-2\cosh\varphi_0}}=
\frac{F\left.\left(\arccos\left[{\sinh\frac{\varphi_0}{2}}\left/
\sinh\frac{\varphi(x)}{2}\right.\right]\right|1/\cosh^2 \frac{\varphi_0}{2}\right)}
{\cosh \frac{\varphi_0}{2}},  \label{eqn:implic}
\qquad
\eea
whose solution is written in terms of 
the  
midplane electrostatic potential $\varphi_0<0$ and $F(\varphi|m)=$\linebreak
 $\int_0^\varphi {\dd\theta}/{\sqrt{1-m
\sin^2\theta}}$ is the incomplete elliptic integral of the first 
kind \cite{erdelyi,abramowitz,byrd,gradshteyn}.
Applying the boundary conditions~(\ref{eqn:boundarycond}) yields  
\bea
\frac{2}{\lambda}
&=&\sqrt{2\cosh\varphi_L-2\cosh\varphi_0}, \\
l&=&\int\limits_{\varphi_L}^{\varphi_0} \frac{\dd\varphi}
{\sqrt{2\cosh\varphi-2\cosh\varphi_0}}=
\frac{F\left.\left(\arccos\left[{\sinh\frac{\varphi_0}{2}}\left/
\sinh\frac{\varphi_L}{2}\right.\right]\right|1/\cosh^2\frac{\varphi_0}{2}\right)}
{\cosh \frac{\varphi_0}{2}},
\qquad
\eea
where we defined the two dimensionless
distances $\lambda\equiv\kb\Lambda$ and $l\equiv\kb L$, 
and $\varphi_L<\varphi_0<0$ is the  surface electrostatic 
potential at the charged planes.
Introducing the variable
\be
t \equiv  \sinh^2 \frac{\varphi_0}2,
\ee 
which \textit{a posteriori} will be identified with half of the 
(dimensionless)
 osmotic-pressure difference between the interplane region and the
salt reservoir --- cf. Eq.~(\ref{eqn:pbmidplane}) --- 
the two boundary conditions can be 
combined into the eigenvalue equation, 
\be
l\sqrt{1+t}= F\left[\arctan\left(\frac1{\lambda\sqrt{t}}\right)
\left|\frac{1}{1+t}\right.\right],
\qquad\mbox{ or }\qquad
\lambda \sqrt{t}= 
\cs\left(l  \sqrt{1+t}\left|\frac1{1+t}\right.\right),
\label{eqn:eigenvalue}
\ee
where $\cs(u|m)=\cn(u|m)/\sn(u|m)$ 
is the ratio of the cosine-amplitude and sine-amplitude 
Jacobi elliptic functions \cite{erdelyi,abramowitz,byrd,gradshteyn}.
The explicit exact solution of the nonlinear PB problem can then be written 
as
\be
\varphi(x)=-2\arcsinh\left[\sqrt{t}\left/ 
\cn\left(\kb |x|\sqrt{1+t}\left|\frac1{1+t}\right.\right)\right.\right],
\qquad\quad |x|\leq L. \label{eqn:explic}
\ee
It should be remarked that the exact solution to the nonlinear PB problem 
may be cast in several equivalent forms. Verwey and
Overbeek~\cite{verwey}, quoted by Hunter~\cite{hunter}, 
gave an alternative form for the implicit solution~(\ref{eqn:implic}), 
\bea
\kb|x|&=& l +2 \mathrm{e}^{-\varphi_0/2} 
\left[F\left(\arcsin\mathrm{e}^{-{\left(\varphi_L-\varphi_0\right)}/{2}}
\left|\,\mathrm{e}^{-2\varphi_0}\right.\right)-
F\left(\arcsin\mathrm{e}^{-{\left[\varphi(x)-\varphi_0\right]}/{2}}\left|\,
\mathrm{e}^{-2\varphi_0}\right.\right)\right]\qquad\nonumber\\
&=& 2 \mathrm{e}^{-\varphi_0/2} 
\left[K\left(\mathrm{e}^{-2\varphi_0}\right)-
F\left(\arcsin\mathrm{e}^{-{\left[\varphi(x)-\varphi_0\right]}/{2}}\left|\,
\mathrm{e}^{-2\varphi_0}\right.\right)\right],
\eea
where $K(m)=F(\pi/2\,|\,m)$ is the complete elliptic integral of the first
kind \cite{erdelyi,abramowitz,byrd,gradshteyn}, 
while Behrens and Borkovec's version~\cite{behrens}  
to the explicit solution~(\ref{eqn:explic}) reads 
\be
\varphi(x)=\varphi_0 + 2 \ln{\mathrm{cd}}
\left(\mathrm{e}^{-\varphi_0/2}{\kb|x|}/{2}
\,\left|\,\mathrm{e}^{2\varphi_0}\right.\right),
\ee
where $\mathrm{cd}(u|m)$ is the cd Jacobi elliptic 
function \cite{erdelyi,abramowitz,byrd,gradshteyn}.
However, none of these previous works presented the explicit 
expression for the nonlinear PB semi-grand-canonical 
potential $\Omega\equiv{\mathit{\Omega}}[n_\pm(x)]_\mathrm{equil}$, 
which can be extracted from Ref.~[\citen{brushes}] by neglecting 
the electrostatic and elastic contributions arising from the
polyelectrolyte brushes.  

The dimensionless \textit{excess} \cite{footnote3}
 semi-grand-canonical 
potential per unit 
area,
\be
\omega(\lambda,l)\equiv
\frac{\kb}{2\nb}
\left[\frac{\beta \Omega(\Lambda,L)}{A}+2\nb L\right], 
\ee
may be evaluated inserting the exact 
nonlinear solution~(\ref{eqn:explic}) into the 
semi-grand-canonical functional~(\ref{eqn:omegapb})
and performing the integrations.
After some tedious algebra, we obtain
\bea
\omega(\lambda,l) &=& 
\frac2{\lambda}\left(2\coth\frac{\varphi_L}2 - \varphi_L\right)+
4E\left(\arccos\left[{\sinh\frac{\varphi_0}{2}}\left/
\sinh\frac{\varphi_L}{2}\right.\right]\left|\,
1/\cosh^2 \frac{\varphi_0}{2}\right.\right)\cosh\frac{\varphi_0}{2}\ \nonumber\\
&&-2F\left(\arccos\left[{\sinh\frac{\varphi_0}{2}}\left/
\sinh\frac{\varphi_L}{2}\right.\right]\left|\,
1/\cosh^2 \frac{\varphi_0}{2}\right.\right)
\sinh\frac{\varphi_0}{2}\tanh\frac{\varphi_0}{2} 
, \\
\omega(\lambda,\infty)&=& 
-\frac2{\lambda}\,\varphi_\infty -8\sinh^2\frac{\varphi_\infty}{4},
\eea
where $E(\varphi|m)=\int_0^\varphi {\dd\theta} 
{\sqrt{1-m\sin^2\theta}}$ is the incomplete 
elliptic integral of the second kind \cite{erdelyi,abramowitz,byrd,gradshteyn},
$\varphi_\infty<0$ is the reduced electrostatic surface potential at the 
charged plane at infinite separation and
$\omega(\lambda,\infty)$ represents the nonlinear excess 
self-energy of the system.  
Using the relations
\be
\cosh\varphi_L=1+2t +\frac{2}{\lambda^2},\qquad\qquad
\cosh\varphi_\infty=1+\frac{2}{\lambda^2},
\ee
and the fact that $\varphi_L<0$ and $\varphi_\infty<0$, 
the excess semi-grand-canonical potential $\omega$ may be cast in a
simpler form, 
\bea
\omega(\lambda,l)&=&
\frac{2}{\lambda} \arccosh\left(1+2t
+\frac{2}{\lambda^2}\right) -\frac{4}{\lambda}
\sqrt{\frac{1+\lambda^2(1+t)}{1+\lambda^2t}}
+4\sqrt{1+t}\,E\left[\arctan\left(\frac1{\lambda\sqrt{t}}\right)\left|
\frac1{1+t}\right.\right] -{2t l}, 
\label{eqn:pbomega} \qquad\\
\omega(\lambda,\infty)&=& 
\frac{2}{\lambda}\arccosh\left(1+\frac2{\lambda^2}\right)
+4\left(1-\frac1{\lambda}\sqrt{1+\lambda^2}\,\right).\qquad
\eea
As already pointed out in the introduction, besides
its intrinsic relevance, the semi-grand-canonical potential 
for the planar case has also 
an important application: by using the Derjaguin 
approximation \cite{derjaguin,white,israelachvili}, valid when the 
range of interactions \cite{footnote4} (of order $\kb^{-1}$) 
and the separation distance $2L$ between the two curved surfaces 
are much smaller than their curvature radius $a$, it is possible 
to determine their normal force (per unit area) ${\cal F}$ 
at separation $2L$ by
\be
\frac{1}{\kb a}{\cal F}(L)= \frac{k_B T}{\lb} 
\left[\omega(\lambda,l)-\omega(\lambda,\infty)\right]. 
\ee 

Returning to the planar case, the osmotic-pressure difference $\Delta P$  
between the interplane region and the infinite salt reservoir 
can be written in terms of the 
midplane reduced electrostatic potential $\varphi_0$, 
\bea
\Pi&\equiv&
\frac{\beta\Delta P}{2\nb} =
-\frac{1}{2\nb}\,\frac{\dd}{\dd L}
\left[\frac{\beta\Omega(\Lambda,L)}{A}+2\nb L
\right]_{\sigma,\lb,\mu_\pm}= 
-\frac{\dd\omega(\lambda,l)}{\dd l}=
\frac{n_+(0)+n_-(0)-2\nb}{2\nb}
\qquad\nonumber\\&=& 
2 \sinh^2 \frac{\varphi_0}{2}
=2t, \label{eqn:pbmidplane}
\eea
which may be obtained by taking the formal derivative of 
$\omega$ with respect to $l$, similarly as 
performed in Appendix~A of Ref.~[\citen{sphere}]. 
Eq.~(\ref{eqn:pbmidplane}) is a mean-field version \cite{marcus}
of the boundary-density theorem, which states that
the osmotic pressure is simply given by
the sum of the microionic densities at the midplane 
(WS-cell boundary). 
This simple relation does not hold beyond the mean-field 
level because of finite ionic-size effects 
and the presence of microionic correlations 
between particles located in the different semi-spaces 
separated by the midplane --- even though it still does 
for \textit{one} charged plane with the electrolyte confined by 
a \textit{neutral} midplane \cite{wennerstrom}. 
We restrict ourselves, however,  
to the nonlinear mean-field result~(\ref{eqn:pbmidplane}),
which clearly predicts that the 
osmotic-pressure difference $\Pi$ \textit{is always positive.}

We should remark that the nonlinear osmotic-pressure
difference~(\ref{eqn:pbmidplane}) may also be written in a 
gauge-invariant form using the gauge-invariant equilibrium 
density profiles~(\ref{eqn:npmgauge}),
\be 
\Pi= 
\left[\sqrt{\left(\frac{2}{\lambda l}\right)^2 +
{\alpha_+\alpha_-}}+\frac{2}{\lambda l}
\right]\frac{\mathrm{e}^{\left\langle\psi\right\rangle-\psi_0}}{2\alpha_+}+ 
\left[\sqrt{\left(\frac{2}{\lambda l}\right)^2 +
{\alpha_+\alpha_-}}-\frac{2}{\lambda l}
\right]\frac{\mathrm{e}^{-\left\langle\psi\right\rangle+\psi_0}}{2\alpha_-}
-1 , \label{eqn:pbinvariant}
\ee
where $\psi_0$ is the \textit{arbitrary} midplane 
electrostatic potential in the gauge-invariant formulation.
It does not need necessarily to coincide with the midplane 
electrostatic potential $\varphi_0\equiv 2\arcsinh\sqrt{\Pi/2}$ 
in the standard gauge $\mu_\mathrm{el}\equiv 0$.
The gauge-invariant form~(\ref{eqn:pbinvariant})
 of the nonlinear osmotic-pressure difference 
will be useful later, at the end of Section~\ref{section:3},
when establishing 
a connection between its quadratic expansion about the 
average potential $\left\langle\psi\right\rangle$ and its 
linearized counterpart~(\ref{eqn:pidh2}). 

\section{Linearization scheme\label{section:3}}

To obtain the linearized semi-grand-canonical functional 
$\mathit{\Omega}_\mathrm{DH}[n_\pm(x)]$
we truncate the expansion of the PB nonlinear semi-grand-canonical functional 
$\mathit{\Omega}[n_\pm(x)]$, Eq.~(\ref{eqn:omegapb}), 
up to second order in the differences $n_\pm(x)-
\left\langle n_\pm(x)\right\rangle$,
where $\left\langle n_\pm(x)\right\rangle\equiv (1/L)\int_0^L 
\dd x\,n_\pm(x)$ are the (\textit{a priori} unknown) average 
densities,
\bea
\frac{\beta{\mathit{\Omega}}_\mathrm{DH}}{A}
&=&\frac1{8\pi\lb}\int\limits_0^L\dd x\left[\frac{\dd\psi(x)}{\dd x}\right]^2
+L \sum_{i=\pm} \left\langle n_i\right\rangle
\left[\ln\frac{\left\langle n_i\right\rangle}{\nb}-1\right]
+\sum_{i=\pm} \left[\left\langle n_i\right\rangle
\ln{\left\langle n_i\right\rangle}\right] 
\int\limits_0^L \dd x \left[\frac{n_i(x)}
{\left\langle n_i\right\rangle}-1\right] \nonumber\\
&&+\frac12\sum_{i=\pm}\left\langle n_i\right\rangle
\int\limits_0^L \dd x \left[\frac{n_i(x)}
{\left\langle n_i\right\rangle}-1\right]^2.
\label{eqn:omegadh}
\eea
After minimization of the functional 
$\mathit{\Omega}_\mathrm{DH}[n_\pm(x)]$ with respect to the 
profiles $n_\pm(x)$
under the overall electroneutrality
constraint~(\ref{eqn:neut_planes}), 
$\left[{\delta}/{\delta n_\pm(x)}\right]
\left[\mathit{\Omega}_\mathrm{DH}/A-
\mu_\mathrm{el}\int \dd x\,\rho(x) \right]=0$
  --- analogously as performed for the spherical case in 
Appendix~F of Ref.~[\citen{sphere}] --- 
we obtain the self-consistent linearized averaged densities,
\be
\bar n_\pm\equiv \left\langle n_\pm(x)\right\rangle_1 =
\frac{\sqrt{\nc^2+(2\nb)^2}\pm\nc}{2},
\label{eqn:donnandensities}
\ee
and the linearized equilibrium density profiles, 
\be
n_\pm(x) =\bar n_\pm\left[1\pm 
\left\langle\psi(x)\right\rangle_1 \mp \psi(x) \right], \label{eqn:dhprofiles}
\ee
where the notation $\left\langle \cdots \right\rangle_1$ 
emphasizes the fact that the self-consistent averaged 
densities~(\ref{eqn:donnandensities}) 
were obtained within a \textit{linearized} approximation. Henceforth
we will omit the subscript `1' in order to simplify the notation.
We should remark that the profile independence of the  
self-consistent linearized averaged
densities given by~(\ref{eqn:donnandensities}) can only be 
verified \textit{after} the minimization procedure. 
When performing the functional minimization of the linearized functional
$\mathit{\Omega}_\mathrm{DH}[n_\pm(x)]$, one  must take
the profile dependence of the average expansion 
densities $\left\langle n_\pm(x)\right\rangle$ into account, in
addition to the charge-neutrality constraint~(\ref{eqn:neut_planes}). 
Although similar quadratic expansions about the zero-th order Donnan 
densities for the planar case were
already proposed by Trizac and 
Hansen~\cite{trizachansen}, they focused their study 
on finite-size effects and did not investigate the 
consequences of the linearization in detail.
Deserno and von 
Gr\"unberg \cite{deserno} considered the general $d$-dimensional
problem in a fixed-gauge formulation, 
interpreting these self-consistent linearized 
averaged densities in terms of an optimal
linearization point, $\bar{n}_\pm=
\nb\mathrm{e}^{\mp\bar\psi_\mathrm{opt}}$.

The linearized expansion densities~(\ref{eqn:donnandensities}), 
which correspond to the state-independent zero-th order Donnan densities,
represent the infinite-temperature $(\lb=0)$
limit of the gauge-invariant forms of the equilibrium 
density profiles~(\ref{eqn:npmgauge}) and \textit{do not 
coincide} with the  exact nonlinear averages, 
\be
\left\langle n_\pm(x) \right\rangle =
\frac{\sqrt{\nc^2 +
(2\nb)^2 \alpha_+\alpha_-}\pm \nc}2 
= \frac{\sqrt{\nc^2 +
(2\nb)^2 \mathrm{e}^{\left\langle\delta_2(x)\right\rangle} 
+{\mathrsfs O}\left[\left\langle\delta_3(x)\right\rangle\right]}\pm
\nc}2 , \label{eqn:pbaverages}
\ee
because of the nonvanishing quadratic and higher-order $(\nu\geq 2)$ 
contributions of the electro\-static-potential deviations,
\be
\delta_\nu(x)\equiv 
\left[\left\langle \psi \right\rangle -\psi(x) \right]^\nu. 
\label{eqn:deltanu}
\ee
In Appendix~\ref{appendix:a} we compare the 
uniform self-consistent linearized 
expansion densities~(\ref{eqn:donnandensities}) with the exact 
nonlinear averages~(\ref{eqn:pbaverages}).
Justification of the neglect of the quadratic and higher-order 
contributions under linearized theory, which was done for the spherical
case --- but is trivially generalized for the planar case --- 
is found in~Appendix~F of Ref.~[\citen{sphere}].
However, we should mention that --- although \textit{internal} 
(within the semi-grand-canonical ensemble) self-consistency 
under linearization 
is achieved by using the uniform expansion
densities~(\ref{eqn:donnandensities}) ---  
\textit{global} self-consistency
(preserving the exact nature of the 
Legendre transformation between the 
semi-grand-canonical and canonical ensembles)
requires also the inclusion of the quadratic contribution 
of the averages~(\ref{eqn:pbaverages}), as discussed in detail
in Appendix G
of Ref.~[\citen{sphere}]. The linearized equations including 
these quadratic contributions to the expansion densities 
are presented in Appendix~\ref{appendix:d}, where it is 
shown that their inclusion do not improve the agreement between 
the linearized and the full nonlinear equations. 

Inserting the linearized equilibrium density 
profiles~(\ref{eqn:dhprofiles}) into the 
Poisson equation~(\ref{eqn:poisson}) yields the 
Debye-H\"uckel-like (DH) equation \cite{footnote5},
\bea
\frac{\dd^2\psi(x)}{\dd x^2}=\kappa^2 \left[\psi(x)-
\left\langle \psi(x)\right\rangle -\eta \right] 
+ \frac{2}{\Lambda}\left[\delta(x+L) + \delta(x-L)\right],
\label{eqn:dhlike}
\eea
where the parameter 
\be
\eta\equiv\frac{\bar n_+ -\bar n_-}
{\bar n_+ +\bar n_-}=\frac{\nc}{\sqrt{\nc^2+(2\nb)^2}},  
\ee
measures the relative importance of the counterions to the ionic
strength in the interplane region $|x|\leq L$,
\be
I\equiv \frac12 \left(\bar{n}_++\bar{n}_-\right) = 
\frac12{\sqrt{\nc^2+(2\nb)^2}} =
\frac{\nc}{2\eta}=\frac{\nb}{\sqrt{1-\eta^2}}.
\ee
The (effective) Debye screening length 
in the interplane region $\kappa^{-1}$ satisfies
\be
\kappa^2 =8\pi\lb I= \frac{\kb^2}{\sqrt{1-\eta^2}}> \kb^2,
\ee
showing that screening 
is enhanced compared to the salt reservoir. 

The gauge-invariant linearized electrostatic potential satisfying 
the DH-like equation~(\ref{eqn:dhlike}) subject to 
the boundary conditions~(\ref{eqn:boundarycond}) can be readily obtained, 
\bea
\psi(x)&=&\left\langle \psi(x)\right\rangle +
\eta \left(1-\kappa L\,\frac{\cosh\kappa x}{\sinh \kappa L}\right),
\label{eqn:psilinear}
\eea
where the average electrostatic potential for \textit{an arbitrary}
electrostatic surface potential $\psi_L$ is given by 
\be
\left\langle \psi(x) \right\rangle 
=\psi_L+
\eta\kappa L {\mathrsfs L}\left(\kappa L \right) ,
\ee
in terms of the Langevin function, 
\bea
{\mathrsfs L}(x) &\equiv& \coth x-\frac1{x}.
\eea

The linearized semi-grand-canonical potential, 
${\Omega}_\mathrm{DH}\equiv\mathit{\Omega}_\mathrm{DH}
[n_\pm(x)]_\mathrm{equil}$, is obtaining by 
inserting the equilibrium density 
profiles~(\ref{eqn:dhprofiles}) --- recalling 
that the self-consistent linearized averages, 
$\left\langle n_\pm(x) \right\rangle=\bar{n}_\pm$, 
are given by Eq.~(\ref{eqn:donnandensities}) --- 
and the DH-like solution~(\ref{eqn:psilinear}) into the 
linearized semi-grand-canonical functional 
$\mathit{\Omega}_\mathrm{DH}[n_\pm(x)]$, Eq.~(\ref{eqn:omegadh}).
After performing the integrations, 
we may cast the 
linearized semi-grand-canonical 
potential (per unit area) in the form
\bea
\frac{\beta\Omega_\mathrm{DH}}{A}&=& \sigma 
\left(\arctanh\eta-\frac{1}{\eta}\right)+\frac12{\eta\sigma
\kappa L}{\mathrsfs L}\left(\kappa L\right),
\eea
which yields the dimensionless excess linearized 
semi-grand-canonical potential per unit 
area,
\be
\omega_\mathrm{DH}(\lambda,l)\equiv
\frac{\kb}{2\nb}
\left[\frac{\beta \Omega_\mathrm{DH}(\Lambda,L)}{A}+2\nb L\right]
= \frac2{\lambda}\left[\arctanh\eta-
\frac{1}{\eta}
+\frac12{\eta k  l}{\mathrsfs L}
\left(k l\right) \right]+l, \quad
\label{eqn:omegadhplane}
\ee
written in terms of the dimensionless 
lengths
\be
k^{-1}\equiv \kb \kappa^{-1},\qquad\qquad 
l\equiv\kb L, \qquad\qquad
\lambda \equiv \kb\Lambda. \qquad 
\ee
With these definitions we obtain the 
linearized self-energy 
$\omega_\mathrm{DH}(\lambda,l\to\infty)=2/\lambda^2$
and 
\be 
\eta=\eta(\lambda,l)=
\frac1{\sqrt{1+(\lambda l/2)^2}}, \qquad\quad
 k^2=  k^2(\lambda,l)
= \sqrt{1+ [2/(\lambda l)]^2}=
\frac1{\sqrt{1-\eta^2(\lambda,l)}} .\quad  
\ee

The dimensionless linearized osmotic-pressure difference 
is then given by 
\bea
\Pi_\mathrm{DH}&\equiv&\frac{\beta\Delta P_\mathrm{DH}}{2\nb}= 
-\frac{1}{2\nb}\,\frac{\dd}{\dd L}
\left[\frac{\beta\Omega_\mathrm{DH}(\Lambda,L)}{A}+2\nb L
\right]_{\sigma,\lb,\mu_\pm} =
-\frac{\dd\omega_\mathrm{DH}(\lambda,l)}{\dd l} 
\nonumber\\
&=& k^2 \left\{1+
\eta^2 \left(1-\frac34\eta^2\right) k l\,{\mathrsfs L}\left( k l\right)
+ \frac12{\eta^2}
\left(1-\frac12{\eta^2}\right) \left( k l\right)^2
\left[{\mathrsfs L}^2\left( k l\right)-1\right]
\right\}-1, \qquad \label{eqn:pidh1}
\eea
where we have made use of the total derivative,
\be
\frac{\dd}{\dd l} = \frac{\partial}{\partial l}+ 
 \frac{\dd\eta }{\dd l}\,\frac{\partial}{\partial\eta}+
\frac{\dd k}{\dd l}\,\frac{\partial}{\partial k}= 
\frac{\partial}{\partial l}
-\frac{\eta}{l}\left(1-\eta^2\right) \,\frac{\partial}{\partial\eta}
-\frac{ k\eta^2}{2 l}\,\frac{\partial}{\partial k},
\qquad
\ee
and the derivative of the Langevin function,
\be
{\mathrsfs L}'(x) = 1-\frac2x\,{\mathrsfs L}(x)-{\mathrsfs L}^2(x).
\ee

Alternatively, the same expression for the linearized 
osmotic-pressure difference 
may be obtained by performing a
quadratic expansion of the gauge-invariant form of the 
nonlinear PB osmotic pressure~(\ref{eqn:pbinvariant}) 
--- similarly as obtained for the spherical case 
in Appendix~E of Ref.~[\citen{sphere}],
\be
\Pi_\mathrm{DH}= k^2 \left[1+
\eta\delta_1(0)+
\frac12\delta_2(0) 
-\frac12{\eta^2}\left\langle \delta_2(x) \right\rangle
\right]-1,\qquad  \label{eqn:pidh2}
\ee
where the $\nu$-th order 
electrostatic-potential differences~(\ref{eqn:deltanu}) read 
\be
\delta_\nu(x)= 
\eta^\nu \left(\kappa L\,\frac{\cosh\kappa x}{\sinh \kappa L}-1\right)^\nu. 
\ee
In the next section we will investigate the properties of the 
linearized osmotic-pressure difference defined by 
Eqs.~(\ref{eqn:pidh1})~or~(\ref{eqn:pidh2})
and compare it with its 
exact nonlinear counterpart~(\ref{eqn:pbmidplane}).

\section{Comparison of the exact nonlinear and the linearized 
equations of state\label{section:4}}

As already pointed out in the literature \cite{grunberg,deserno}, 
the linearized osmotic-pressure difference $\Pi_\mathrm{DH}$ 
defined by Eqs.~(\ref{eqn:pidh1})~or~(\ref{eqn:pidh2})
yields artifacts in the low-temperature, large-separation
or high-surface charge limits. In contradiction to the exact nonlinear 
result~(\ref{eqn:pbmidplane}), which 
predicts that the osmotic-pressure difference is always 
positive, $\Pi>0$, the linearized version
$\Pi_\mathrm{DH}$ becomes negative in the above mentioned limits.
In an attempt to define the osmotic pressure in a
 linearized framework, Deserno and von Gr\"unberg \cite{grunberg}
 introduced an additional (alternative) definition,
 $\Pi_1$, cf.~Eq.~(43) of Ref.~[\citen{deserno}],
that does not have the shortcoming of displaying any  
 instabilities. On the other hand, we will show later that  
their partially unstable osmotic-pressure definition, 
cf.~Eq.~(44) of Ref.~[\citen{deserno}],
coincides with the linearized version~(\ref{eqn:pidh1}) 
obtained in the previous section, $\Pi_2\equiv\Pi_\mathrm{DH}$.
Their general formulas, for the planar case $(d=1)$, 
need to be taken in the formal limit
of vanishing volume fraction, $\phi\equiv a/l \to 0$ --- with
$a>0$ being some arbitrary (reduced) length --- which yields 
\bea
\Pi_1&\equiv& \frac{\beta P_1}{2\nb}-1=\cosh\bar\psi_\mathrm{opt}-
\frac{\sinh^2\bar\psi_\mathrm{opt}}{2\cosh\bar\psi_\mathrm{opt}}
\left[1-
\lim_{a\to 0}\left(\frac{1-a/l}{{\cal D}\sqrt{a/l}}\right)^2\right] -1 
\nonumber\\&=& 
\frac{1}{2k^2}\left(k^2-1\right)^2+
\frac12\left(\frac{\eta k^2l}{\sinh kl}\right)^2 \geq 0,
\label{eqn:pidesernogrun}\\
\Pi_2&\equiv&\frac{\beta P_2}{2\nb}-1= \frac{\beta P_1}{2\nb}- 
\frac{\sinh^4\bar\psi_\mathrm{opt}}{2\cosh^3\bar\psi_\mathrm{opt}}
\lim_{a\to 0} \left\{\frac{1-a/l}{2a/l}\left[\frac1{{\cal D}^2}+
ka \frac{\cal E}{\cal D}+
k^2 a^2 \left(1-\frac{{\cal E}^2}{{\cal D}^2}
\right)\right]-1\right\}\nonumber\\
&=& \Pi_1- \frac14{k^2\eta^4}
\left( \frac{k^2 l^2}{\sinh^2 kl}+ k l \coth kl -2\right) ,\\
{\cal D}&=& K_{1/2}(ka)I_{1/2}(k l) -
K_{1/2}(k l)I_{1/2}(ka), \\ 
{\cal E}&=& K_{-1/2}(ka)I_{1/2}(k l) + 
K_{1/2}(k l)I_{-1/2}(ka), 
\eea
where $\left\{I_\nu,K_\nu\right\}$ are 
the modified Bessel functions \cite{abramowitz} of the first and the 
second kind, respectively, and 
$\bar\psi_\mathrm{opt}$ is the optimal linearization point, 
satisfying the relations
\be
\tanh \bar\psi_\mathrm{opt}=-\eta, \qquad\quad
\cosh\bar\psi_\mathrm{opt}=\left(\frac{\kappa}{\kb}\right)^2=k^2,
\quad\qquad \sinh\bar\psi_\mathrm{opt}=-\frac{\nc}{2\nb}=-\frac{2}{\lambda l}.
\ee
In accordance with Eqs.~(23) and (26) of Ref.~[\citen{deserno}],
the two osmotic-pressure definitions 
can be recast in a simpler formal form, 
\bea
\Pi_1 &=& k^2 \left[1+
\eta\delta_1(0)+
\frac12\delta_2(0) \right]-1
, \label{pi1formal}\qquad \\
\Pi_2&=& k^2 \left[1+
\eta\delta_1(0)+
\frac12\delta_2(0)- \frac12{\eta^2}
\left\langle\delta_2(x)\right\rangle\right]-1,\qquad
\eea
from which one can see that the second osmotic-pressure definition
coincides with the linearized 
osmotic-pressure difference~(\ref{eqn:pidh1}) obtained in the last section, 
$\Pi_2\equiv\Pi_\mathrm{DH}$, while the first one 
$\Pi_1$ differs from Eq.~(\ref{eqn:pidh2}) 
by an omitted quadratic term. 
Analogously to the spherical case \cite{sphere}, 
the term that distinguishes the two distinct osmotic-pressure definitions 
originates from the volume dependence of the optimal linearization
point $\bar\psi_\mathrm{opt}$, as pointed out by Deserno and 
von Gr\"unberg \cite{deserno}. 

From its asymptotic expansions to be given next
and its formal expression~(\ref{pi1formal}), 
we see that $\Pi_1$, although \textit{fully thermodynamically
stable} --- related to its positiveness, 
 cf.~Eq.~(\ref{eqn:pidesernogrun}) --- 
is \textit{inconsistent} with a quadratic expansion of the 
gauge-invariant nonlinear 
PB pressure~(\ref{eqn:pbinvariant}), because of the omitted last quadratic 
term of~(\ref{eqn:pidh2}). Furthermore, we will show next that the 
\textit{consistent} --- although \textit{partially unstable} --- 
linearized osmotic-pressure difference $\Pi_2$ presents 
indeed a better agreement with the nonlinear osmotic pressure $\Pi$ in the 
weak-coupling and counterionic ideal-gas limits, when 
the underlying assumptions of the linearization are fulfilled.
Therefore, although the alternative 
$\Pi_1$ displays the \textit{fortuitous} advantage of 
preserving the positiveness of the exact nonlinear 
pressure $\Pi$, its derivation has no justification in our
approach based on the minimization of the linearized
semi-grand-canonical functional 
 $\mathit{\Omega_\mathrm{DH}}[n_\pm(x)]$. 
Moreover, the partially unstable $\Pi_2$
corresponds indeed to the negative \textit{total derivative} of the 
linearized semi-grand-canonical potential $\omega_\mathrm{DH}$ with 
respect to the planes separation $l$, which we thus believe to be the 
consistent and correct definition of the osmotic pressure. 

Let us now perform an explicit comparison between asymptotic expressions 
of $\Pi$, the nonlinear osmotic pressure \cite{footnote6} --- obtained in 
Appendix~\ref{appendix:b} --- and of the 
two corresponding linearized versions, $\Pi_1$ and $\Pi_2$, 
 for the distinct regimes listed below. 
\smallskip

\noindent$\bullet$ Weak-coupling or zero-th order Donnan $(\lb\to 0)$ limit: 
$l\to 0$, $\lambda\to\infty$, 
but finite product $\lambda l$
\bea
\Pi&=&
k^2 \left[1-\frac{\eta^2}6{k^2 l^2} 
-\frac{\eta^2}{90}{\left(\eta^2-3\right) k^4 l^4} 
+\frac{\eta^2}{945}{\left(3\eta^2-5\right) k^6 l^6}
-\frac{\eta^2}{113400}{\left(7\eta^6+18\eta^4+51\eta^2-84\right)k^8 l^8}
\right.\nonumber\\&&\left.\vphantom{\frac{\eta^2}6{k^2 l^2}}
+{\mathrsfs O}\left({k^{10} l^{10}}\right)\right]-1,\qquad
\label{eqn:pblb} \\
\Pi_1&=& 
k^2 \left[1-\frac{\eta^2}6{k^2 l^2} 
+\frac{\eta^2}{30}{k^4 l^4} -\frac{\eta^2}{189}{k^6 l^6} 
+\frac{\eta^2}{1350}{k^8 l^8} 
+{\mathrsfs O}\left({k^{10} l^{10}}\right)\right]-1,\qquad \\
\Pi_2&=&
k^2 \left[1-\frac{\eta^2}6{k^2 l^2} 
-\frac{\eta^2}{90}{\left(\eta^2-3\right) k^4 l^4} 
+\frac{\eta^2}{945}{\left(2\eta^2-5\right) k^6 l^6}
-\frac{\eta^2}{9450}{\left(3\eta^2-7\right) k^8 l^8}
+{\mathrsfs O}\left({k^{10} l^{10}}\right)\right]-1,\qquad 
\eea
\smallskip

\noindent$\bullet$ Counterionic ideal-gas limit:
$l\to 0$ and finite $\lambda$
\bea
\Pi&=&
\frac{2}{\lambda l}\left\{1-\frac{l}{3\lambda}+ 
\left(\frac{4}{45}+\frac{\lambda^4}{8}\right)
\left(\frac{l}{\lambda}\right)^2 
-\frac{16}{945}\left(\frac{l}{\lambda}\right)^3
+{\mathrsfs O}\left[ \left(l/\lambda\right)^4
\right]\right\}-1,\qquad \label{eqn:pbsmallsep}\\
\Pi_1&=&
\frac{2}{\lambda l}\left\{1-\frac{l}{3\lambda}+ 
\left(\frac{2}{15}+\frac{\lambda^4}{8}\right)
\left(\frac{l}{\lambda}\right)^2 
-\frac{8}{189}\left(\frac{l}{\lambda}\right)^3
+{\mathrsfs O}\left[ \left(l/\lambda\right)^4
\right]\right\}-1 ,\qquad \\
\Pi_2&=&
\frac{2}{\lambda l}\left\{1-\frac{l}{3\lambda}+ 
\left(\frac{4}{45}+\frac{\lambda^4}{8}\right)
\left(\frac{l}{\lambda}\right)^2 
-\frac{8}{315}\left(\frac{l}{\lambda}\right)^3
+{\mathrsfs O}\left[ \left(l/\lambda\right)^4
\right]\right\}-1 ,\qquad \label{eqn:dhsmallsep}
\eea
\smallskip

\noindent$\bullet$ Gouy-Chapman or high-surface charge
 limit:  $l\to 0$ and $\lambda/l\to 0$
\bea
\Pi&=& 
\frac12\left(\frac{\pi}{l}\right)^2\left\{1-\frac{2\lambda}{l}
+{\mathrsfs O}\left[\left({\lambda}/{l}\right)^3\right] \right\}
-1 + {\mathrsfs O}\left(l^2\right), \label{eqn:pblambda} \\
\Pi_1&=&
\frac{2}{\lambda^2\sinh^2\sqrt{{2 l}/
{\lambda}}}+\frac{1}{\lambda l}-
\frac{l^2 \sqrt{{2 l}/{\lambda}}\coth\sqrt{{2 l}/{\lambda}}}
{4\sinh^2\sqrt{{2 l}/{\lambda}}} -1
+{\mathrsfs O}\left(\sqrt{{\lambda}/
{l}}\,\right), \qquad \\
\Pi_2&=&
\frac{1}{\lambda^2\sinh^2\sqrt{{2 l}/
{\lambda}}} - 
\frac{\coth\sqrt{{2 l}/
{\lambda}}}{\lambda^2\sqrt{2l/\lambda}}+
\frac{2}{\lambda l}+{\mathrsfs O}\left(\sqrt{{l}/
{\lambda}}\,\right),\label{eqn:lambdatozero}
\eea
\smallskip

\noindent$\bullet$ Large-separation limit:
$l\to\infty$ and finite $\lambda$ 
\bea
\Pi&=& 
\frac{32 \mathrm{e}^{-2 l}}
{\left(\lambda+\sqrt{1+\lambda^2}\,\right)^2}
\left\{1-8\left[l-1+\frac{\lambda(1-\lambda^2)}{\sqrt{1+\lambda^2}}\right]
\frac{\mathrm{e}^{-2l}}
{\left(\lambda+\sqrt{1+\lambda^2}\,\right)^{2}} +
{\mathrsfs O}\left(l^2\mathrm{e}^{-4 l}\right)\right\},
\label{eqn:pbhs} \\
\Pi_1&=&
\frac{1}{\sinh^2 l}\left[
\frac{2}{\lambda^2}-\frac{4\coth l}{\lambda^4 l}
-\frac{2\left(1-3\coth^2l\right)}{\lambda^6 l^2}
+\frac{2\left(8+9\lambda^2-12\coth^2l\right)\coth l}{3\lambda^8 l^3}
\right]+\frac{2\left[1+
{\mathrsfs O}\left(\mathrm{e}^{-2 l}\right)\right]}{\lambda^4 l^4}
\nonumber\\&&
+{\mathrsfs O}\left({l}^{-5}\right), \\
\Pi_2&=&
\frac{1}{\sinh^2 l}\left[
\frac{2}{\lambda^2}-\frac{4\coth l}{\lambda^4 l}
-\frac{2\left(1+2\lambda^2-3\coth^2l\right)}{\lambda^6 l^2}
\right]-\frac{4\left[1+
{\mathrsfs O}\left(\mathrm{e}^{-2 l}\right)\right]}{\lambda^4 l^3}
+{\mathrsfs O}\left({l}^{-4}\right). \quad 
\label{eqn:phitozero}
\eea

Looking at~Eqs.~(\ref{eqn:lambdatozero})~and~(\ref{eqn:phitozero})
one may see why the linearized osmotic-pressure difference 
$\Pi_2$ becomes negative at the Gouy-Chapman and 
large-separation limits. In the Gouy-Chapman limit
the leading term is given by the 
${\mathrsfs O}\left({\lambda}^{-3/2}\right)$, which is
negative and overcomes the exponentially decaying 
${\mathrsfs O}\left({\lambda}^{-2}\right)$ term.
The leading term of the large-separation limit 
is given by the ${\mathrsfs O}\left({l}^{-3}\right)$
contribution, which is negative and overcomes the three exponentially 
decaying lowest-order terms. 
In the full nonlinear solution, however, 
all algebraically decaying terms cancel in a nontrivial way, and 
eventually only an exponentially (positive) decaying behavior is
predicted. 
Note that both linearized versions, 
$\Pi_1$ and $\Pi_2$, show asymptotic behaviours 
that disagree strongly from the 
nonlinear osmotic-pressure difference $\Pi$.
This clearly indicates that both linearized osmotic-pressure 
definitions are meaningless in these limits and so is the 
positiveness of $\Pi_1$. 

We see that in the weak-coupling limit the self-consistent linearized 
osmotic pressure $\Pi_2$ and its nonlinear counterpart $\Pi$
agree up to the ${\mathrsfs O}\left({l}^{4}\right)$
terms, confirming the validity of the linearization when its underlying
assumptions are fulfilled. The same occurs for the 
 counterionic ideal-gas limit 
up to  the ${\mathrsfs O}\left({l}\right)$ terms. In both cases 
the fully stable $\Pi_1$ has a worse agreement, one order lower
than the partially unstable $\Pi_2$. 
However, in the large-separation limit, the two linearized and the 
nonlinear expressions disagree even qualitatively: the linearized 
asymptotics is algebraic (negative for $\Pi_2$,
positive for $\Pi_1$), 
whereas the nonlinear is exponential (and positive).
On the other hand, although in the Gouy-Chapman limit all 
asymptotics are 
algebraic, in the linearized case the power-law is 
$\propto -{l}^{-1/2}$ for $\Pi_2$ and 
$\propto {l}^{-1}$ for $\Pi_1$, 
both in disagreement with the nonlinear asymptotics 
$\propto {l}^{-2}$.
The failure of the linearization scheme 
should not be at all surprising, 
because it is supposed to be valid in the weak-coupling 
$(\lb\to 0)$ and counterionic ideal-gas limit, but not in 
the opposite, large-separation  $(l\to\infty)$ or 
high-surface charge, Gouy-Chapman, $(\lambda\to 0)$ limits.
Therefore, any results obtained in a linearized 
framework outside the weak-coupling and the counterionic ideal-gas 
limits should be taken with caution.
See also our comments in the concluding remarks of the
preceding paper \cite{sphere} on the gas-liquid-like phase separation 
in dilute deionized aqueous suspensions of charged colloidal 
particles.

In order to show the accuracy of the self-consistent  linearized 
osmotic-pressure difference $\Pi_\mathrm{DH}$, 
Eq.~(\ref{eqn:pidh1}), and 
the region where the linearization scheme breaks down,
we plotted in Figures~\ref{figure:1}~and~\ref{figure:2} 
the locii of constant errors between the exact 
nonlinear PB osmotic-pressure difference and the corresponding 
linearized version, 
measured by the logarithmic deviations 
\bea
\delta\Pi_\mathrm{DH}&\equiv& \left|\ln\Pi_\mathrm{DH}(\lambda,l)-
\ln\Pi(\lambda,l)\right|. \label{eqn:pidhdeviation}
\eea
We have chosen a logarithmic measure for the deviations because 
$\Pi$ varies in a range of several orders of magnitude. For small 
deviations, this definition leads to the relative errors,
\bea
\delta\Pi_\mathrm{DH}&\approx&\left| 
\frac{\Pi_\mathrm{DH}(\lambda,l)-\Pi(\lambda,l)}
{\Pi(\lambda,l)}\right|.
\eea  
Analogously, we may define the logarithmic deviation from PB 
of the linearized semi-grand-canonical potential, 
\bea
\delta\omega_\mathrm{DH}&\equiv& \left| \ln\left[
\omega_\mathrm{DH}(\lambda,l)-\omega(\lambda,\infty)
\right]-
\ln\left[\omega(\lambda,l)-
\omega(\lambda,\infty)\right]\right| ,
\eea
which is always smaller than $\delta\Pi_\mathrm{DH}$ (not shown).
Therefore the linearized semi-grand-canonical potential, 
Eq.~(\ref{eqn:omegadhplane}),
and the linearized osmotic-pressure difference, 
Eq.~(\ref{eqn:pidh1}), describe well the corresponding 
nonlinear equations in the limit $\lambda/l\gg 1$ and $l\ll 1$.
Because the nonlinear
theory always predicts repulsion, the attractive osmotic-pressure 
region --- shown in gray in 
 Figures~\ref{figure:1}~and~\ref{figure:2} --- 
 is clearly an artifact of the linearization. When plotted 
on the $\Lambda/L\times 
(\kb L)^{-1}$ plane, the $\Pi_\mathrm{DH}=0$ line reaches 
at $\kb L\to 0$ the asymptotic value
$\xi_0=\Lambda/L=0.123863965\cdots$, which is 
given by the solution of the transcendental equation
\be
2\xi_0+
\sqrt{\frac{\xi_0}{2}}{\mathrsfs L}\left(\sqrt{\frac{2}
{\xi_0}}\,\right)+ 
{\mathrsfs L}^2\left(\sqrt{\frac{2}{\xi_0}}\,\right)
=1. \label{eqn:xi}
\ee

To obtain the full nonlinear PB osmotic-pressure difference 
 $\Pi$, one needs
to numerically solve the 
transcendental equation~(\ref{eqn:eigenvalue})
involving elliptic functions or elliptic integrals. 
Although the asymptotic expansions of the nonlinear 
$\Pi$ represented 
by Eqs.~(\ref{eqn:pblb}),~(\ref{eqn:pbsmallsep}),%
~(\ref{eqn:pblambda})~and (\ref{eqn:pbhs})
allow an explicit analytical comparison in the distinct regimes
with their linearized versions, 
they are not very useful for numerical evaluation. 
In Appendix \ref{appendix:c} we derive extended 
expansions of the nonlinear PB semi-grand-canonical potential
$\omega$, Eq.~(\ref{eqn:pbomega}), 
and of the PB osmotic-pressure difference $\Pi$, 
 Eq.~(\ref{eqn:pbmidplane}), 
that involve only elementary functions and are suitable
for numerical implementation. 
These extend the numerical accuracy of the above mentioned  
asymptotic expansions of the full nonlinear $\Pi$ and 
are complementary to the linearized equations,
$\omega_\mathrm{DH}$, Eq.~(\ref{eqn:omegadhplane}), 
and $\Pi_\mathrm{DH}$, Eq.~(\ref{eqn:pidh1}), 
providing an excellent approximation in the 
regions where the linearization scheme breaks down.
In Figures~\ref{figure:1}~and~\ref{figure:2} we also 
present their corresponding logarithmic deviations from the 
exact PB result, which, similarly 
to~(\ref{eqn:pidhdeviation}), are defined by 
\bea
\delta\Pi_\mathrm{GC}&\equiv& \left| \ln\Pi_\mathrm{GC}(\lambda,l)-
\ln\Pi(\lambda,l)\right|,\\
\delta\Pi_\mathrm{LS}&\equiv& \left|\ln\Pi_\mathrm{LS}(\lambda,l)-
\ln\Pi(\lambda,l)\right|,
\eea
where $\Pi_\mathrm{GC}$ and $\Pi_\mathrm{LS}$, 
given explicitly in Appendix~\ref{appendix:c},
are the osmotic-pressure
differences in the extended Gouy-Chapman and large-separation 
limits, respectively.

\section{Concluding remarks\label{section:5}}

The classical problem of two infinite uniformly charged planes 
in electrochemical equilibrium with an infinite salt-reservoir is
exactly solved at the mean-field nonlinear level, as well as 
by a linearization scheme consistent with quadratic
expansions of the nonlinear semi-grand-canonical functional.
By using gauge-invariant forms of the electrostatic potential,
we have shown that the linearized osmotic pressure corresponds 
to a quadratic expansion of the corresponding nonlinear version. 

As already pointed out in the literature \cite{deserno},
it is shown that the self-consistent linearized osmotic pressure
leads to artifacts in the large-separation and the Gouy-Chapman 
(high-surface charge) limits, predicting there negative 
osmotic-pressure differences. 
Although it is possible to define an alternative 
linearized osmotic pressure
that it is fully stable based on the \textit{partial 
derivative} of the linearized semi-grand-canonical potential 
with respect to the separation distance \cite{deserno}, 
its stability is shown to be a fortuitous result.
In fact explicit comparison of the exact 
nonlinear osmotic pressure
and the two linearized versions shows that the 
linearized self-consistent osmotic pressure, though partially unstable,  
presents a better agreement 
with the PB results in the weak-coupling and counterion 
 ideal-gas limits, 
where the linearization can be applied. 
However, not surprisingly, in the region where the linearization
breaks down none of both proposed linearized 
osmotic pressures give quantitatively
correct results. 

To avoid confusion we should stress at this point the exactness of
the PB nonlinear solution at the mean-field level and discuss
its range of validity and limitations.  
It is known from numerical simulations of the Primitive 
Model \cite{friedman} (PM)  in the planar geometry 
that sufficiently close and highly 
charged planes in the presence of neutralizing counterions 
attract each other~\cite{guldbrand}, even though 
for realistic charge densities and monovalent ions this is 
not observed at room temperature. In this case the attraction is prevented by 
steric repulsions at the small separations at which it would be
observed neglecting the finite ionic size. 
Because the mean-field PB approximation always predicts repulsion, 
theoretical validation for this attraction (observed in fact at room 
temperature only for multivalent ions) has to be given beyond the 
PB level, e.g., by 
bulk counterion correlations~\cite{stevens,diehlplates}, 
integral-equations theories~\cite{kjellander,cassou},
charge-correlation-induced attractions~\cite{rouzina,shklovskii1}, 
charge-fluctuation-induced attractions~\cite{pincussafran,lau,ha1},
electrolytic depletion-induced attractions~\cite{tamashirodepletion},
discrete solvent-mediated attractions~\cite{burak}, field-theory methods~\cite{netz} etc --- see 
also~Refs.~[\citen{groenberg,podgornik,ha,arenzon1,solis,%
shklovskii2,arenzon2,diehl,jurgen,desernoaxel}] 
 for mechanisms of attraction between like-charged rods. 
On the other hand, in the strong-coupling limit the linearization 
of the WS-cell mean-field PB equation, as discussed in this work,  
 \textit{does predict} attraction. 
However, here the mechanism of attraction 
is related to mathematical artifacts 
of the linearization itself and does not correspond to a real 
physical effect. The fact that this prediction
is in agreement with the theories beyond the mean-field level is purely
\textit{accidental} and is intrinsically connected with 
the \textit{inadequacy} (meaning incorrect application) of 
the PB mean-field approach at the same limit. In other words, a 
\textit{qualitatively} correct result (in this example, attraction) 
may be deceptively anticipated in the strong-coupling limit 
because of the simultaneous application  
of two inadequate approximations, namely, 
the mean-field PB equation and its subsequent 
linearization. 


\section*{Acknowledgments}

M.~N.~T. would like to thank the Alexander von Humboldt-Stiftung
for financial support.

\appendix
\setcounter{equation}{0}
\renewcommand{\thesection}{\Alph{section}}
\renewcommand{\theequation}{\Alph{section}\arabic{equation}}
\section{Exact nonlinear averaged densities\label{appendix:a}}

In this Appendix we will compare the uniform 
expansion densities about which the
linearization is performed ---  the state-independent 
 zero-th order Donnan densities --- 
with the exact nonlinear PB averages.

\begin{figure}[ht]
\begin{center}
\epsfig{file=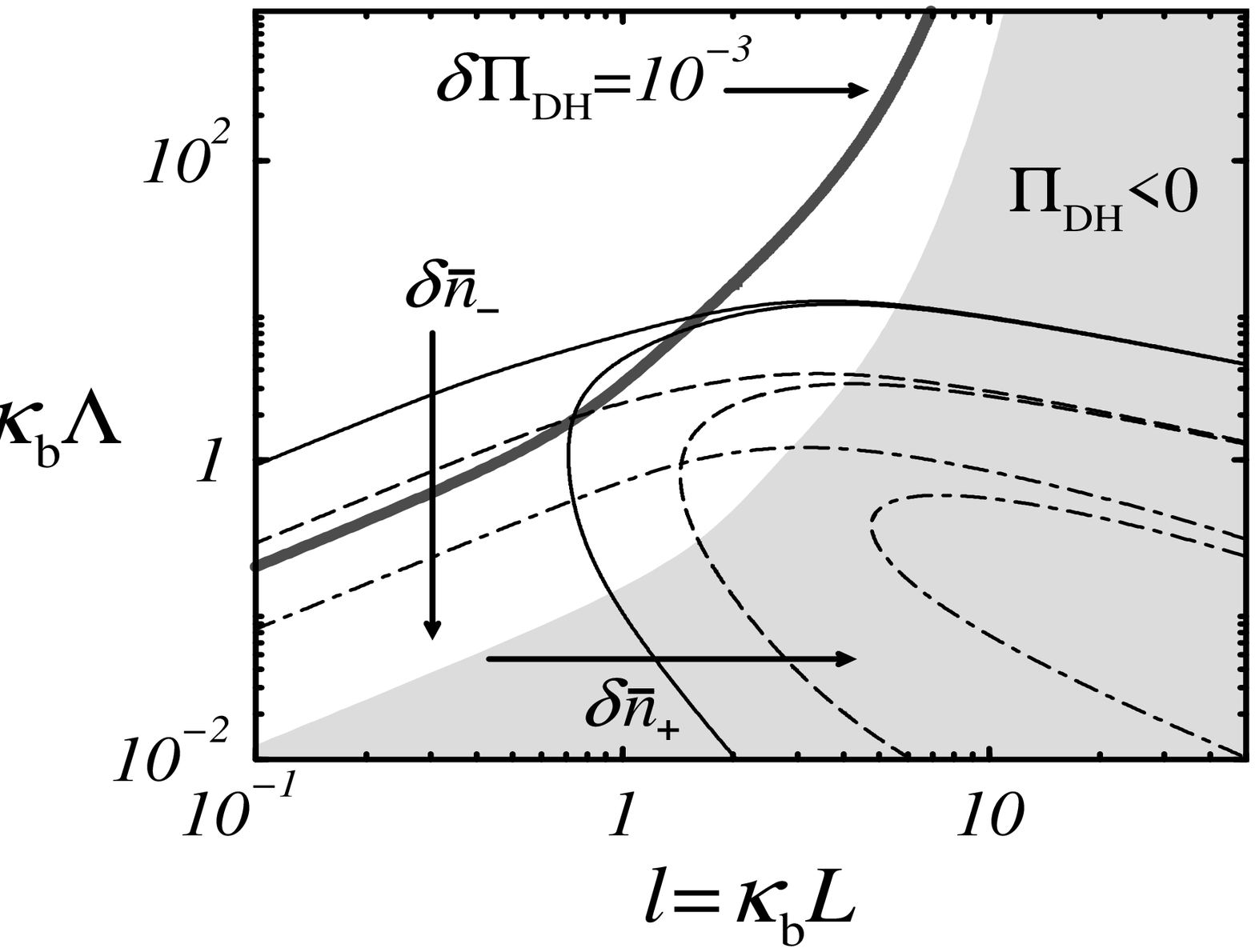,width=0.65\textwidth}
\caption{Deviations from the PB averaged densities 
of the state-independent zero-th order Donnan densities, which 
were used to perform the quadratic expansions of the 
nonlinear functional. 
The arrows indicate the direction of increasing 
logarithmic deviations $\delta\bar{n}_\pm$ 
from the PB results:  $10^{-3}$ (solid lines),
$10^{-2}$ (dashed lines) and $10^{-1}$ (dot-dashed lines).
To allow a comparison with the region where the linearized theory
breaks down, we also plotted the locus 
(dark gray thick line) where the logarithmic deviation
from PB of the linearized osmotic-pressure difference is 
$\delta\Pi_\mathrm{DH}=10^{-3}$. In the 
light gray region the linearized osmotic-pressure difference becomes negative,
$\Pi_\mathrm{DH}<0$. Although there is a close connection 
between this region and the increase of the deviations 
$\delta\bar{n}_\pm$ for high-surface charges ($\lambda\ll 1$), 
for low-surface charges 
($\lambda\gg 1$) and 
large separations  $(l \gg 1)$, the linearized theory still 
predicts a negative linearized osmotic-pressure difference 
 (upper-right region), while the full nonlinear one vanishes 
 exponentially from positive values.  
\label{figure:3}}
\end{center}
\end{figure}

By using the definite integrals
\bea
\int\limits_{\varphi_L}^{\varphi_0}
\frac{\dd\varphi\,\sinh\varphi}{\sqrt{2\cosh\varphi-2\cosh\varphi_0}} &=& 
-{\sqrt{2\cosh\varphi_L-2\cosh\varphi_0}}, \\
\int\limits_{\varphi_L}^{\varphi_0}
\frac{\dd\varphi\,\cosh\varphi}{\sqrt{2\cosh\varphi-2\cosh\varphi_0}} &=&
\frac{\cosh\varphi_0}{\cosh \frac{\varphi_0}{2}}\,
F\left.\left(\arccos\left[{\sinh\frac{\varphi_0}{2}}\left/
\sinh\frac{\varphi_L}{2}\right.\right]\right|1/\cosh^2 \frac{\varphi_0}{2}\right)
 \nonumber\\
&&-2\cosh\frac{\varphi_0}{2}\, 
E\left.\left(\arccos\left[{\sinh\frac{\varphi_0}{2}}\left/
\sinh\frac{\varphi_L}{2}\right.\right]\right|1/\cosh^2
\frac{\varphi_0}{2}\right)\qquad\nonumber\\
&&-\coth\frac{\varphi_L}{2}{\sqrt{2\cosh\varphi_L-2\cosh\varphi_0}},
\eea
it is possible to obtain the exact nonlinear PB averaged densities, 
\bea
\frac{\left\langle n_\pm(x)\right\rangle}{\nb}&=&
\left\langle \mathrm{e}^{\mp\varphi(x)}\right\rangle =
\sqrt{\left(\frac2{\lambda l}\right)^2 + 
\left\langle\mathrm{e}^{\varphi(x)}\right\rangle  
\left\langle \mathrm{e}^{-\varphi(x)}\right\rangle} 
\pm \frac2{\lambda l}\nonumber\\
&=&1+2t \pm \frac2{\lambda l}
-\frac{2}l\sqrt{1+t}\,E\left[\arctan\left(\frac1{\lambda\sqrt{t}}\right)\left|
\frac1{1+t}\right.\right] +
\frac2{\lambda l} \sqrt{\frac{1+\lambda^2(1+t)}{1+\lambda^2 t}}.\qquad
\eea
In Figure~\ref{figure:3} we compare them with the uniform
densities about which the
linearization is performed, the state-independent zero-th order Donnan
densities~(\ref{eqn:donnandensities}), 
\be
\frac{\bar{n}_\pm}{\nb}= 
\sqrt{\left(\frac2{\lambda l}\right)^2+1}\pm \frac2{\lambda l}, 
\ee
by looking at their logarithmic deviations from the 
corresponding exact PB averages, 
\be
\delta\bar{n}_\pm\equiv 
\ln \left\langle{n}_\pm (x)\right\rangle - \ln \bar{n}_\pm.
\ee

\section{Asymptotic expansions of the nonlinear solution\label{appendix:b}}
\setcounter{equation}{0}

In this Appendix we obtain the 
asymptotic expansions of the nonlinear osmotic-pressure 
difference $\Pi$. We have made extensive use of 
Refs.~[\citen{erdelyi,abramowitz,byrd,gradshteyn}] throughout this Appendix.

\subsection{Weak-coupling limit}

To obtain the weak-coupling $(\lb\to 0)$ limit, first note that the
product $\lambda l =2/(\eta k^2)$ does not 
depend on $\lb$. Therefore we multiply both sides of the 
eigenvalue equation by $l$, 
\bea
\lambda l  \sqrt{t}&=& \frac{2\sqrt{t}}{\eta k^2}=
l\cs\left(l  \sqrt{1+t}\left|\frac1{1+t}\right.\right),
\eea
and expand them in powers of $l$, 
assuming that $t=\sum_{k=0}^{\infty} a_{2k}  l^{2k}$. It is then
possible to obtain the coefficients $\{a_{2k}\}$ of the expansion
recursively, leading to~Eq.~(\ref{eqn:pblb}).

\subsection{Counterionic ideal-gas limit}

To obtain the counterion-dominated ideal-gas limit ($l\to 0$ and finite 
$\lambda$), we write 
\bea
\lambda \sqrt{t}&=&
\cs\left(l  \sqrt{1+t}\left|\frac1{1+t}\right.\right),
\eea
and expand the right-hand side in powers of $l$, 
assuming that $t=\sum_{k=-1}^{\infty} a_{k}  l^{k}$. It is then
possible to obtain the coefficients $\{a_{k}\}$ of the expansion
recursively, leading to~Eq.~(\ref{eqn:pbsmallsep}).

\subsection{Gouy-Chapman limit}

To obtain the Gouy-Chapman (high-surface charge) limit 
$(\lambda\to 0)$, note that
\bea
l\sqrt{1+t}&=& 
F\left[\frac{\pi}2-\arctan(\lambda\sqrt{t})\left|
\frac1{1+t}\right.\right]=
K\left(\frac1{1+t}\right)- 
\lambda\sqrt{1+t}+{\mathrsfs O}(\lambda^3t),\qquad
\eea
leading to 
\be
l\left[1+\frac{\lambda}{l} 
+{\mathrsfs O}(\lambda^3t/l)\right]
= \frac{1}{\sqrt{1+t}}\,K\left(\frac1{1+t}\right).
\ee
If we additionally assume that $l\to 0$, with $\lambda/l\to 0$, 
and with the help of the expansion of the complete elliptic 
integral $K(m)$ about $m=0$,
\be
K(m)= \frac{\pi}{2}
\left[1+\frac{m}4+ {\mathrsfs O}(m^2)\right],
\ee
we are lead to the asymptotic solution, 
\be
t=\left(\frac{\pi}{2 l}\right)^2\left\{1-\frac{2\lambda}{l}
+{\mathrsfs O}\left[\left({\lambda}/{l}\right)^3\right] \right\}
-\frac12 + {\mathrsfs O}\left(l^2\right),
\ee
which gives the asymptotic osmotic-pressure 
difference in the Gouy-Chapman limit, Eq.~(\ref{eqn:pblambda}).
Evaluation of higher-order terms of the leading contribution 
${\mathrsfs O}(l^{-2})$ of the 
expansion would give 
\bea
t&=&
\left(\frac{\pi}{2 l}\right)^2\left\{1-\frac{2\lambda}{l}
+ 3 \left(\frac{\lambda}{l}\right)^2 
+\left(\frac{\pi^2}{6}-4\right)\left(\frac{\lambda}{l}\right)^3
-5\left(\frac{\pi^2}{6}-1\right)\left(\frac{\lambda}{l}\right)^4 
-\left(\frac{\pi^4}{40}-\frac{5\pi^2}{2}+6\right)
\left(\frac{\lambda}{l}\right)^5
\right.\nonumber\\&&\left. \vphantom{\left(\frac{\lambda}{l}\right)^5}
+{\mathrsfs O}\left[\left({\lambda}/{l}\right)^6\right] \right\}
-\frac12 + {\mathrsfs O}\left(l^2\right).
\eea

\subsection{Large-separation limit \label{subsection:lslimit}}

To obtain the large-separation $(l\to\infty \mbox{ and finite }
\lambda)$ 
asymptotics, first note that
\bea
l\sqrt{1+t}&=& F\left[\arccotg\left(\lambda\sqrt{t}\,\right)\left|
\frac1{1+t}\right.\right] = 
F\left[\frac{\pi}2-\arctan\left(\lambda\sqrt{t}\,\right)\left|
\frac1{1+t}\right.\right] 
\nonumber\\&=& 
K\left(\frac1{1+t}\right)-
F\left[\arctan\left(\lambda\sqrt{1+t}\,\right)\left|
\frac1{1+t}\right.\right] \nonumber\\
&=& K\left(\frac1{1+t}\right)-
F\left(\arctan\lambda\,|\,1\right)+{\mathrsfs O}(t)
= K\left(\frac1{1+t}\right)-
\ln\left(\lambda+\sqrt{1+\lambda^2}\,\right)
+{\mathrsfs O}(t). \qquad \eea
Using the expansion of the complete elliptic integral $K(m)$ about $m=1$,
\be
K(m)= \ln 4-\frac12\ln(1-m) +  {\mathrsfs O}(1-m),
\ee
we are led to the asymptotic solution
\be
t=\frac{16 \mathrm{e}^{-2 l}}
{\left(\lambda+\sqrt{1+\lambda^2}\,\right)^2} + 
{\mathrsfs O}\left(l\mathrm{e}^{-4 l}\right),
\ee
which yields the leading term of the asymptotic 
osmotic-pressure difference in the 
large-separation limit, Eq.~(\ref{eqn:pbhs}).
By computing higher-order terms, 
it is possible to write an asymptotic power series,
\be
t=\frac{16 \mathrm{e}^{-2 l}}
{\left(\lambda+\sqrt{1+\lambda^2}\,\right)^2}
\left[1+
\sum_{\nu=1}^{\infty}
\frac{\mathrm{e}^{-2\nu l}\,t_\nu(\lambda,l)}
{\left(\lambda+\sqrt{1+\lambda^2}\,\right)^{2\nu}} \right] 
\label{eqn:pils},
\ee
where the coefficients $t_\nu(\lambda,l)={\cal{O}}(l^\nu)$ of the 
three leading terms read
\bea
t_1(\lambda,l)&=&
-8\left[l-1+\frac{\lambda(1-\lambda^2)}{\sqrt{1+\lambda^2}}\right] , \\ 
t_2(\lambda,l)&=&96l^2
-152 l\left[1-
\frac{24\lambda(1-\lambda^2)}{19\sqrt{1+\lambda^2}}\right]+
\frac{4}{1 + \lambda^2}
\left(11 + 35\lambda^2 - 48\lambda^4 + 24\lambda^6\right)
\nonumber\\&&
-\frac{8\lambda}{\left(1 + \lambda^2\right)^{3/2}}
\left(19- 12\lambda^2 - 17\lambda^4 + 6\lambda^6\right), \\
t_3(\lambda,l)&=&-\frac{4096l^3}3
+4096 l^2 \left[\frac{11}{16}-
\frac{\lambda(1-\lambda^2)}{\sqrt{1+\lambda^2}}\right]
+ \frac{512\lambda l}{(1+\lambda^2)^{3/2}}
\left(11 - 6\lambda^2 - 10\lambda^4 + 3\lambda^6\right) \nonumber\\
&& -\frac{32l}{(1+\lambda^2)^2}  
\left(47 + 222\lambda^2 - 81\lambda^4 - 128\lambda^6
  +128\lambda^8\right) \nonumber \\
&& -\frac{32\lambda}{(1+\lambda^2)^{5/2}}
\left(47+15\lambda^2-159\lambda^4+17\lambda^6+
104\lambda^8-56\lambda^{10}\right)\nonumber\\
&&+ \frac{64}{(1+\lambda^2)^3}
\left(3 + 53\lambda^2 - 39\lambda^4 - 77\lambda^6 + 
72\lambda^8 + 36\lambda^{10} - 24\lambda^{12}\right) . 
\eea
Although this asymptotic series works pretty well for large 
separations $(l\gg 1)$, in the crossover region $(l\approx 1)$ 
it leads to oscillating pressures.
In Appendix~\ref{appendix:c} we obtain extended expansions
of the nonlinear equations in the large-separation limit 
that do not have this disadvantage in the crossover region. 

\section{Extended expansions of the nonlinear solution\label{appendix:c}}
\setcounter{equation}{0}

In this Appendix we obtain 
extended expansions of the nonlinear
semi-grand-canonical potential and of the nonlinear 
osmotic-pressure difference 
that are valid in the region where the linearization breaks down. 
Again, We have made extensive use of 
Refs.~[\citen{erdelyi,abramowitz,byrd,gradshteyn}] throughout this Appendix.

\subsection{Extended Gouy-Chapman limit}

In the previous Appendix, both the counterionic ideal-gas (finite $\lambda$) 
as well as the Gouy-Chapman $(\lambda\to 0)$ asymptotics 
were obtained in the small-separation 
 $(l\to 0)$ limit. In fact, for any ratio $\xi=\lambda/l$ 
the summation over the $\lambda/l$ series for the leading 
terms up to ${\mathrsfs O}(l^{2})$ may be performed exactly, yielding
\be
t=\left(\frac{y}{l}\right)^2-\frac12 +
\frac{3\left(1+\xi+\xi^2 y^2\right)
\left(1+\xi^2 y^2\right)+2\xi^3 y^2}
{32 y^2\left(1+\xi+\xi^2 y^2\right)
\left(1+\xi^2 y^2\right)}\, l^2 +
{\mathrsfs O}\left(l^4\right),
\ee
where $y=y(\xi)$ is the solution 
of the transcendental equation 
\be
\xi y \tan y = 1.  \label{eqn:yxi}
\ee
This general expression yields the leading term ${\mathrsfs O}(l^{-2})$
of the counterionic ideal-gas 
 (finite $\lambda$, when $y\to\sqrt{l /\lambda}\to 0$) 
as well as the Gouy-Chapman 
(high-surface charge, when $\lambda\to 0, y\to \pi/2)$ asymptotics  
as special cases,
\bea
y^2&=&\frac{l}{\lambda}\left\{ 1-
\frac{l}{3\lambda}
+\frac4{45}\left(\frac{l}{\lambda}\right)^2
-\frac{16}{945}\left(\frac{l}{\lambda}\right)^3
+\frac{16}{14175} \left(\frac{l}{\lambda}\right)^4
+\frac{64}{93555} \left(\frac{l}{\lambda}\right)^5
+{\mathrsfs O}\left[\left({l}/{\lambda}\right)^6\right]\right\}
\nonumber\\
&=& \left(\frac{\pi}{2}\right)^2
\left\{1-\frac{2\lambda}{l}
+ 3 \left(\frac{\lambda}{l}\right)^2 
+\left(\frac{\pi^2}{6}-4\right)\left(\frac{\lambda}{l}\right)^3
-5\left(\frac{\pi^2}{6}-1\right)\left(\frac{\lambda}{l}\right)^4 
-\left(\frac{\pi^4}{40}-\frac{5\pi^2}{2}+6\right)
\left(\frac{\lambda}{l}\right)^5 
\right.\nonumber\\
&&\left. \vphantom{\left(\frac{\lambda}{l}\right)^5} 
+{\mathrsfs O}\left[\left({\lambda}/{l}\right)^6\right] \right\}.
\eea 
The excess semi-grand-canonical potential $\omega$ may be 
obtained by integration of the osmotic-pressure difference $2t$, 
leading to 
\bea
\omega(\lambda,l) 
&=& \frac{2y^2}{l}-\frac{4}{\lambda} 
\left[1+\ln\left(\frac{\lambda}{2}\sin y\right)\right]+l
-\frac12\cot y\left(\sin^2 y +
\frac{l}{\lambda}\right)\left(\frac{l}{2y}\right)^3
+ {\mathrsfs O}\left(l^5\right).\quad
\eea
While the third and fourth terms are the 
leading corrections due to the presence of salt,
the two first terms can be related to half of the
\textit{exact nonlinear} 
Helmholtz free energy of two charged infinite planes 
in the presence of neutralizing 
counterions only (salt-free Gouy-Chapman case),
\be
\frac{\beta F}{A}=\frac1{4\pi\lb} 
\left\{\frac{2y^2}{L}-\frac{4}{\Lambda} 
\left[1+\ln\left({\Lambda}\sin y\right) + \frac12 
\ln\left(\frac{2\pi\lb}{\zeta^3}\right)\right] \right\},
\ee
where $y$ is the solution of the transcendental equation
$y \tan y = L/\Lambda$. 

We define the extended Gouy-Chapman limit
by truncating the above expansions, neglecting thus
higher-order terms, 
\bea
\omega_\mathrm{GC}(\lambda,l) 
&\equiv& \frac{2y^2}{l}-\frac{4}{\lambda} 
\left[1+\ln\left(\frac{\lambda}{2}\sin y\right)\right]+l
-\frac12\cot y\left(\sin^2 y +
\frac{l}{\lambda}\right)\left(\frac{l}{2y}\right)^3, \qquad\\
\Pi_\mathrm{GC}(\lambda,l) &\equiv&
2\left(\frac{y}{l}\right)^2-1 +
\frac{3\left(1+\xi+\xi^2 y^2\right)
\left(1+\xi^2 y^2\right)+2\xi^3 y^2}
{16 y^2\left(1+\xi+\xi^2 y^2\right)
\left(1+\xi^2 y^2\right)}\, l^2,
\eea
where $y=y(\xi)$ is the solution 
of the transcendental equation~(\ref{eqn:yxi}).

\subsection{Extended large-separation limit}

The large-separation osmotic-pressure  
asymptotics~(\ref{eqn:pils}) obtained 
in Appendix~\ref{subsection:lslimit}
displays oscillations in the 
crossover $(l\approx 1)$ region. 
Because we want to match
the linearized DH-like, the extended Gouy-Chapman 
and the large-separation asymptotic expressions at the 
crossover region, we need to find an extended expansion 
that does not display this shortcoming.
In fact the pressure oscillations are avoided if one 
uses instead the implicit form $l=l(\lambda,m)$, 
which is obtained 
by expanding the eigenvalue equation,
\bea
l(\lambda,m)&=&\sqrt{m}\,K\left(m\right)-
\sqrt{m}\,F\left[\left.\arctan\left(\lambda/\sqrt{m}\,\right)\,
\right|m\right] ,
\eea
in powers of $(1-m)\equiv t/(1+t)$.
Accurate results in the crossover region, which will 
cover almost the whole $(l\times\lambda)$  parameter space 
with logarithmic pressure 
deviations from the exact PB within 0.1\%,  are obtained by 
using fourth-order expansions of the elliptic integrals \cite{footnote7}
about $m=1$,
\bea
K(m)&=& \ln 4-\frac12\ln(1 - m)
+ \frac14(1 - m)\left[\ln 4 -1 - \frac12\ln(1-m)\right] \nonumber\\
&+&\!\!\frac3{128}(1 - m)^2\left[6\ln 4-7 - 3\ln(1-m)\right]
+ \frac5{1536}(1 - m)^3\left[30\ln 4-37 -15\ln(1-m)\right]\nonumber\\
&+&\!\!\frac{35}{196608}(1 - m)^4
\left[420\ln 4-533 - 210\ln(1-m)\right]
+ {\cal O}\left[(1-m)^5\ln(1-m)\right],
\eea
\bea
F\left[\left.\arctan\left({\lambda}/{\sqrt{m}}\,\right)
\right|m\right] &=& F\left(\left.\arctan\lambda\right|m\right) +
\frac{\lambda\left(1-m\right)}{2\left(1 + \lambda^2\right)^{7/2}}
\left\{\left(1 + \lambda^2\right)^3 \right.\nonumber\\
+\frac14\left(1-m\right)\left(1 + \lambda^2\right)^2
\left(3 - 2\lambda^4\right)&+&
\frac1{24}\left(1-m\right)^2\left(1 + \lambda^2\right)
\left(15 + 5\lambda^2 - 10\lambda^4 + 6\lambda^6 + 9\lambda^8\right)
\nonumber\\
+\frac{1}{192}\left(1-m\right)^3
\left(105 + 70\lambda^2-70\lambda^4\!\right.&+&\left.\left.\!28\lambda^6 + 
66\lambda^8 - 72\lambda^{10} - 60\lambda^{12}\right) 
+ {\cal O}\left[(1-m)^4\right]\right\}, \quad\ \  
\eea
\bea
F\left(\arctan\lambda\,|\,m\right) &=& 
F\left(\arctan\lambda|1\right) +
\sum_{n=1}^\infty \frac{(m-1)^n}{n!} 
\left.\frac{\partial^n F\left(\arctan\lambda\,|\,m\right) }
{\partial m^n} \right|_{m=1} \nonumber\\
&=& \ln\left(\lambda+\sqrt{1+\lambda^2}\,\right) +
\frac{1-m}4 
\left[\ln\left(\lambda+\sqrt{1+\lambda^2}\,\right)-
\lambda \sqrt{1+\lambda^2}\,\right]\nonumber\\
+\frac{3\left(1-m\right)^2}{64}&&\!\!\!\!\!\!\left[
3\ln\left(\lambda+\sqrt{1+\lambda^2}\,\right)-
\lambda\sqrt{1+\lambda^2}\left(3-2\lambda^2\right) \right] \nonumber\\
+\frac{5\left(1-m\right)^3}{768}&&\!\!\!\!\!\!\left[
15 \ln\left(\lambda+\sqrt{1+\lambda^2}\,\right)-
\lambda\sqrt{1+\lambda^2}\left(15 - 10\lambda^2 + 8\lambda^4\right)
\right]\nonumber\\
+\frac{35\left(1-m\right)^4}{49152}&&\!\!\!\!\!\!\left[
105 \ln\left(\lambda+\sqrt{1+\lambda^2}\,\right)-
\lambda\sqrt{1+\lambda^2}\left(105-70\lambda^2+
56\lambda^4-48\lambda^6\right) \right] \nonumber\\
+ {\cal O}\left[(1-m)^5\right].&&
\eea
The asymptotic large-separation $(l\to\infty)$
excess semi-grand-canonical 
potential is obtained by integration
of the osmotic-pressure difference, $2t=-\dd\omega/\dd l$,
\bea
\omega(\lambda,l) 
&=& \omega(\lambda,\infty)-
2 \int_0^t \dd \tau\, \tau\,\frac{\dd l}{\dd\tau}
+{\cal O}\left[(1-m)^6\ln(1-m)\right] \nonumber \\
&=&
\omega(\lambda,\infty)-2tl+
2 \int_0^t \dd \tau\, l\left[\lambda,1/({1+\tau})\right] 
+{\cal O}\left[(1-m)^6\ln(1-m)\right]\nonumber\\
&=& \omega(\lambda,\infty)-\frac2{m}\,{(1-m)}\,l(\lambda,m)-
2 \int_1^m \frac{\dd\mu}{\mu^2}\, l(\lambda,\mu)
+{\cal O}\left[(1-m)^6\ln(1-m)\right]. \label{eqn:omegalinfty} \qquad
\eea
We need to evaluate integrals of type
\bea
\int_1^m \frac{\dd\mu}{\mu^{3/2}}\, \left(1-\mu\right)^n 
, \qquad
\int_1^m \frac{\dd\mu}{\mu^{3/2}}\,
\ln\left(1-\mu\right) \left(1-\mu\right)^n , 
\quad\mbox{ for } n=0,\cdots,4,
\eea
which can all be performed analytically.
After some algebraic manipulations, 
the last contribution of Eq.~(\ref{eqn:omegalinfty})
 may be cast in the form 
\bea
2 \int_1^m \frac{\dd\mu}{\mu^2}\, l(\lambda,\mu) &=& 
\frac{315}{32}\left[\arctanh\sqrt{m} + \ln 2- 
\ln\left(\lambda+\sqrt{1+\lambda^2}\,\right)-\frac{19}{56}\right]- 
\frac{\lambda {\cal A}(\lambda)}{3360\left(1
    +\lambda^2\right)^{7/2}} +\frac{{\cal B}(m)}{8192{\sqrt{m}}}\times
\nonumber\\
&\times&\!\!\left[\ln\left(1-m\right)-4\ln2+
2\ln\left(\lambda+\sqrt{1+\lambda^2}\,\right)\right]
+\frac{{\cal C}(m)}{49152{\sqrt{m}}} 
+\frac{\displaystyle\sum_{n=0}^{7}\lambda^{2n+1} {\cal C}_n(m)}
{430080\left(1 +\lambda^2\right)^{7/2}\sqrt{m}} , \qquad\ \ \\
{\cal A}(\lambda)&=&34125 + 24150\lambda^2 - 56210\lambda^4 - 
46480\lambda^6 + 9253\lambda^8 + 
738\lambda^{10}
- 5160\lambda^{12} + 1680\lambda^{14}, \\
{\cal B}(m)&=&25609 + 18404m - 4818m^2 + 1300m^3-175m^4, \\
{\cal C}(m)&=&123743+82792m - 56970m^2 + 16960m^3 - 2365m^4, \\
{\cal C}_0(m)&=&  105\left(14711 + 35356m - 11310m^2 + 
3308m^3 - 465m^4\right), \\
{\cal C}_1(m)&=& 70\left(19795 + 26252m - 1254m^2 - 
868m^3 + 235m^4\right), \\
{\cal C}_2(m)&=& -70\left(31873 + 98756m - 37826m^2 + 
11636m^3 - 1655m^4\right), \\
{\cal C}_3(m)&=& - 560\left(4129 + 7940m - 1730m^2 + 
308m^3 - 23m^4\right), \\
{\cal C}_4(m)&=& 156695 + 1690780m - 938350m^2 + 
322924m^3 - 47665m^4, \\
{\cal C}_5(m)&=& 79030 - 82600m + 186900m^2 - 
111496m^3 + 22630m^4, \\
{\cal C}_6(m)&=& -40\left(4795 + 16940m - 7350m^2 + 
2492m^3 - 365m^4\right), \\
{\cal C}_7(m)&=& 1680\left(35 + 140m - 70m^2 + 
28m^3 - 5m^4\right).
\eea
We may check numerically that the osmotic pressure $2t$ corresponds 
indeed to the derivative of the excess grand potential
$\omega$ with respect to the semi-separation $l$ between the two 
charged surfaces. 

The extended large-separation expressions are obtained by truncating
the above expressions, neglecting thus higher-order contributions. 
Collecting all contributions, the extended large-separation osmotic-pressure 
difference  $\Pi_\mathrm{LS}\equiv{2(1-m)}/m$ is defined 
implicitly by the relation 
\bea
\frac{l(\lambda,m)}{\sqrt{m}}&=&\ln 4-\frac12\ln(1 - m)
+ \frac14(1 - m)\left[\ln 4 -1 - \frac12\ln(1-m)\right] \nonumber\\
&&+\frac3{128}(1 - m)^2\left[6\ln 4-7 - 3\ln(1-m)\right]
+ \frac5{1536}(1 - m)^3\left[30\ln 4-37 -15\ln(1-m)\right]\nonumber\\
&&+\frac{35}{196608}(1 - m)^4
\left[420\ln 4-533 - 210\ln(1-m)\right]-
\frac{\lambda\left(1-m\right)}{2\left(1 + \lambda^2\right)^{7/2}}
\left\{\left(1 + \lambda^2\right)^3 \right.\nonumber\\
&&+\frac14\left(1-m\right)\left(1 + \lambda^2\right)^2
\left(3 - 2\lambda^4\right)+
\frac1{24}\left(1-m\right)^2\left(1 + \lambda^2\right)
\left(15 + 5\lambda^2 - 10\lambda^4 + 6\lambda^6 + 9\lambda^8\right)
\nonumber\\
&&+\left.\frac{1}{192}\left(1-m\right)^3
\left(105 + 70\lambda^2-70\lambda^4+28\lambda^6 + 
66\lambda^8 - 72\lambda^{10} - 60\lambda^{12}\right)\right\} \nonumber\\
&&-\ln\left(\lambda+\sqrt{1+\lambda^2}\,\right) -
\frac{1-m}4 
\left[\ln\left(\lambda+\sqrt{1+\lambda^2}\,\right)-
\lambda \sqrt{1+\lambda^2}\,\right]\nonumber\\
&& -\frac{3\left(1-m\right)^2}{64} \left[
3\ln\left(\lambda+\sqrt{1+\lambda^2}\,\right)-
\lambda\sqrt{1+\lambda^2}\left(3-2\lambda^2\right) \right] \nonumber\\
&& -\frac{5\left(1-m\right)^3}{768} \left[
15 \ln\left(\lambda+\sqrt{1+\lambda^2}\,\right)-
\lambda\sqrt{1+\lambda^2}\left(15 - 10\lambda^2 + 8\lambda^4\right)
\right]\nonumber\\
&&-\frac{35\left(1-m\right)^4}{49152} \left[
105 \ln\left(\lambda+\sqrt{1+\lambda^2}\,\right)-
\lambda\sqrt{1+\lambda^2}\left(105-70\lambda^2+
56\lambda^4-48\lambda^6\right) \right],\qquad
\eea
with the associated semi-grand-canonical potential,  
\be
\omega_\mathrm{LS}(\lambda,l)\equiv
\frac{2}{\lambda}\arccosh\left(1+\frac2{\lambda^2}\right)
+4\left(1-\frac1{\lambda}\sqrt{1+\lambda^2}\,\right)
-\frac2{m}\,{(1-m)}\,l(\lambda,m)-
2 \int_1^m \frac{\dd\mu}{\mu^2}\, l(\lambda,\mu).
\ee

\section{Globally self-consistent linearized equations\label{appendix:d}}
\setcounter{equation}{0}

In this Appendix we show
that the linearized equations that preserve the exact nature of the 
Legendre transformation do not lead to any improvements in the
agreement between the linearized and nonlinear osmotic pressures
in comparison to the linearized versions obtained in Section~\ref{section:3}.  

As discussed in detail in Appendix G of Ref.~[\citen{sphere}],
the Legendre transformation between the semi-grand-canonical
and the canonical descriptions of the system may be rendered 
\textit{exact} if --- instead of using the state-independent zero-th order 
Donnan densities~(\ref{eqn:donnandensities})
--- one uses the quadratic truncation of the nonlinear averages, given
by~Eq.~(\ref{eqn:pbaverages}), as expansion densities to obtain 
the linearized functional. With the inclusion of the 
 quadratic  state-dependent contribution 
$\left\langle \delta_2(x) \right\rangle$ to the average densities,  
we obtain the \textit{globally} 
self-consistent linearized 
semi-grand-canonical potential and linearized osmotic-pressure 
difference,
\bea
\hat\omega_\mathrm{DH}(\lambda,l)&\equiv&
\frac{\kb}{2\nb}
\left[\frac{\beta\hat\Omega_\mathrm{DH}(\Lambda,L)}{A}+2\nb L\right]
=\frac2{\lambda}\left[\arctanh\hat\eta-
\frac{1}{\hat\eta}
+\frac12{\hat\eta\hat{k}l}{\mathrsfs L}
\left(\hat{k}l\right) +
\frac{1}{2\hat\eta}\left\langle \delta_2(x) \right\rangle\right]+l\!
\nonumber\\
&=& \frac2{\lambda}\left\{\arctanh\hat\eta-
\frac{1}{\hat\eta}
+\frac54{\hat\eta\hat{k}l}{\mathrsfs L}
\left(\hat{k}l\right) 
+\frac14{\hat\eta(\hat{k}l)^2}\left[{\mathrsfs L}^2
\left(\hat{k}l\right)-1\right] \right\}+l, \\
\hat\Pi_\mathrm{DH}(\lambda,l)&\equiv&
-\frac{\dd\hat\omega_\mathrm{DH}(\lambda,l)}{\dd l} 
= \hat{k}^2 \left[1+
\hat\eta\delta_1(0)+
\frac12\delta_2(0) 
-\frac12\left\langle \delta_2(x) \right\rangle
\right]-1 \nonumber\\
&=& \hat{k}^2 \left\{1+
\frac14\hat\eta^2 \hat{k}l {\mathrsfs L}\left(\hat{k} l\right)
+\frac14 (\hat\eta\hat{k}l)^2
 \left[{\mathrsfs L}^2\left(\hat{k}l\right)-1  \right] \right\}-1,
\label{eqn:pidhhat}
\eea
where the dimensionless parameters,
\be 
\hat\eta=\frac1{\sqrt{1+(\lambda l/2)^2
\mathrm{e}^{\left\langle \delta_2(x) \right\rangle}}}, 
\qquad\quad
\hat{k}^2=\sqrt{\mathrm{e}^{\left\langle \delta_2(x) \right\rangle}
+ [2/(\lambda l)]^2}=
\frac{\mathrm{e}^{\left\langle \delta_2(x) \right\rangle/2}}
{\sqrt{1-\hat\eta^2}} , \label{eqn:parametric}\quad  
\ee
are given implicitly in terms of the quadratic 
electrostatic-potential deviation,  
\be
\left\langle \delta_2(x) \right\rangle = \frac12 \hat\eta^2\hat{k}l
\left\{3{\mathrsfs L}\left(\hat{k} l\right)
+\hat{k}l \left[{\mathrsfs L}^2\left(\hat{k}l\right)-1\right] 
\right\} . 
\ee
To compute the linearized osmotic-pressure difference $\hat\Pi_\mathrm{DH}$, 
Eq.~(\ref{eqn:pidhhat}), one needs to take  into account
the total derivatives of the parametric forms, 
Eqs.~(\ref{eqn:parametric}),
\bea
\frac{\dd}{\dd l} &=& 
\frac{\partial}{\partial l}+
\frac{\dd\hat\eta }{\dd l}\,\frac{\partial}{\partial\hat\eta}+
\frac{\dd\hat\kappa}{\dd l}\,\frac{\partial}{\partial\hat\kappa}
\nonumber\\&=&\frac{\partial}{\partial l}
-\frac{\hat\eta}l\left(1-\hat\eta^2\right) 
\left(1+\frac{l}{2}\,\frac{\dd \left\langle \delta_2(x)\right\rangle}
{\dd l}\right)\frac{\partial}{\partial\hat\eta}
-\frac{\hat\kappa\hat\eta^2}{2l}
\left[1-\frac{l}{2}\left(\frac{1-\hat\eta^2}{\hat\eta^2}\right)
\frac{\dd \left\langle \delta_2(x)\right\rangle}
{\dd l}\right]\frac{\partial}{\partial\hat\kappa}. \qquad
\eea
In accordance to the infinite-separation linearized self-energy obtained 
in Section~\ref{section:3}, the new version 
 is also given by $\hat\omega_\mathrm{DH}(\lambda,l\to\infty)=2/\lambda^2$.

\begin{figure}
\begin{center}
\epsfig{file=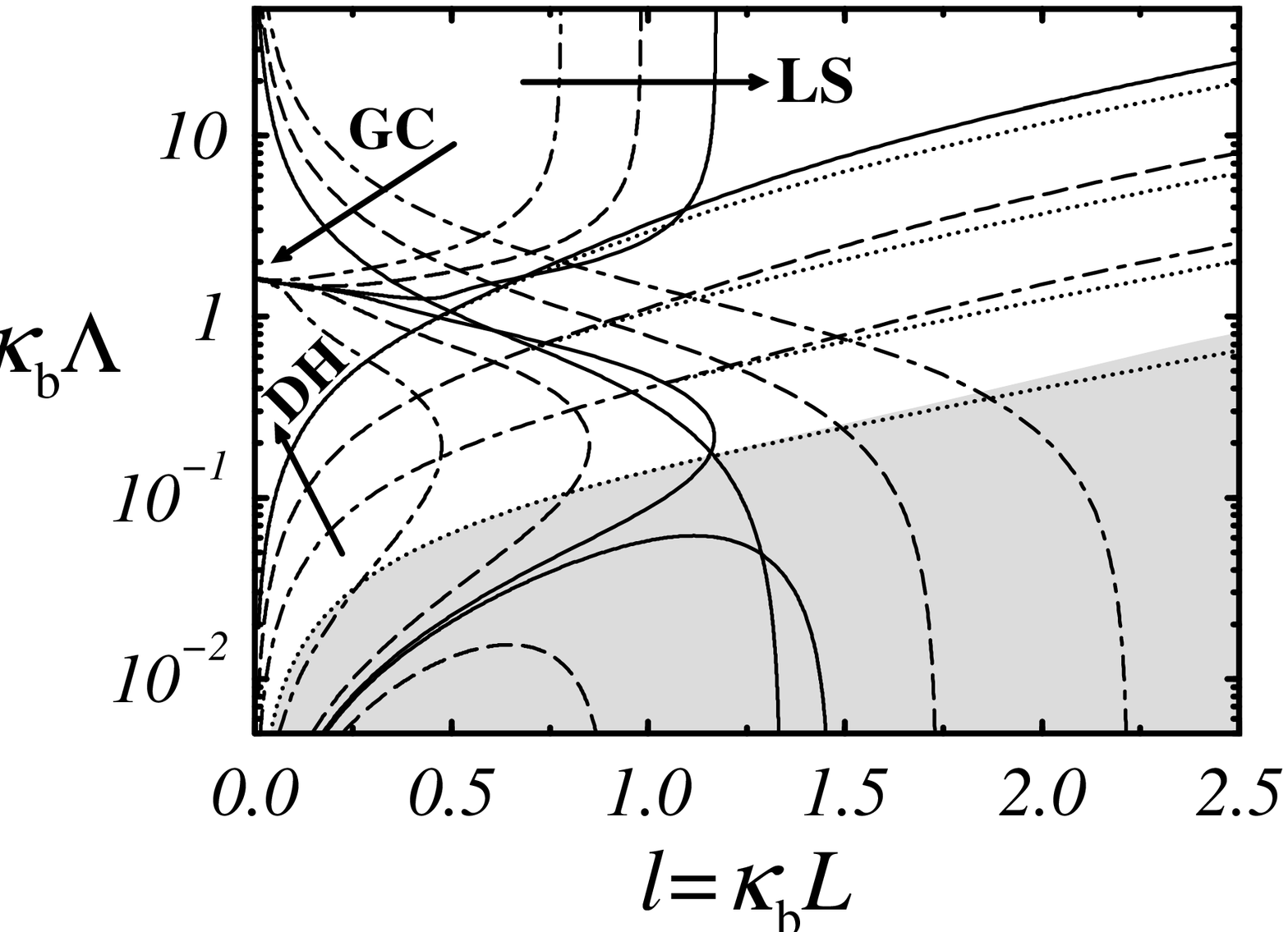,width=0.65\textwidth}
\caption{Logarithmic deviations from the PB of the different 
asymptotic osmotic-pressure differences. DH represents 
the Debye-H\"uckel-like, linearized-functional
expansion about the weak-coupling limit, GC corresponds to the expansion
about the salt-free Gouy-Chapman limit,
and LS denotes the large-separation limit expansion.
The region complementary to LS is splitted into three parts.
In the gray region the linearized osmotic-pressure difference 
$\Pi_\mathrm{DH}$ 
becomes negative. The arrows indicate the direction of decreasing 
logarithmic deviation $\delta\Pi$ from the PB results: $10^{-1}$ (dot-dashed
lines), $10^{-2}$ (dashed lines), $10^{-3}$ (solid lines).
For comparison, we also display (dotted lines) the linearized results 
by including quadratic contributions in the expansion densities, 
as defined by the linearized pressure 
 $\hat\Pi_\mathrm{DH}$, Eq.~(\ref{eqn:pidhhat}).
\label{figure:1}}
\epsfig{file=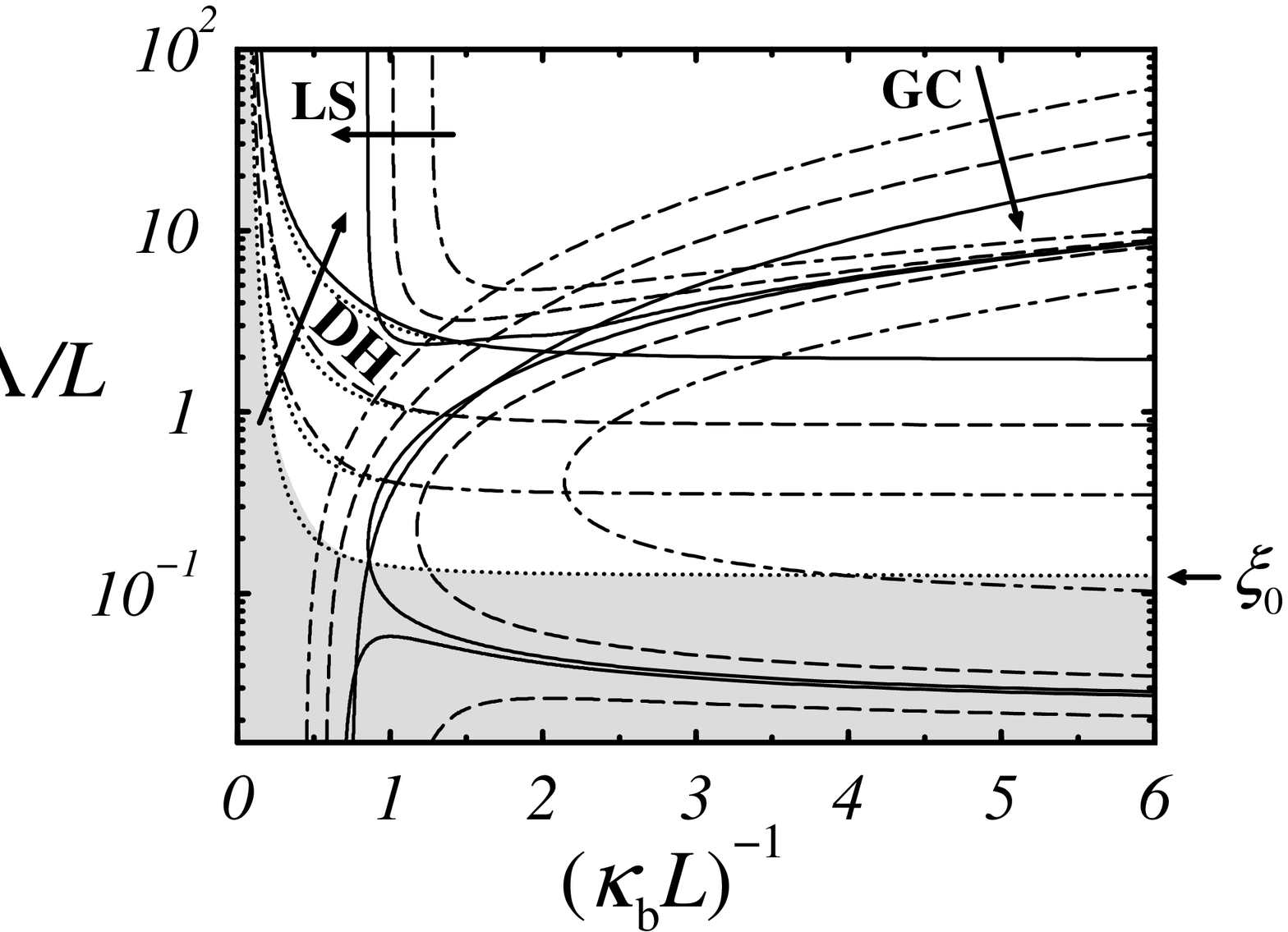,width=0.65\textwidth}
\caption{Same as in Figure~\ref{figure:1}, 
but plotted using different variables.
At $\kb L\to 0$, the $\Pi_\mathrm{DH}=0$ line reaches the asymptotic value
defined by Eq.~(\ref{eqn:xi}), $\xi_0=\Lambda/L=0.123863965\cdots$ 
Compare with Figure~1 from Ref.~[\citen{pincus}].
\label{figure:2}}
\end{center}
\end{figure}

In Figures~\ref{figure:1}~and~\ref{figure:2} 
we compare the two linearized osmotic-pressure 
definitions, $\Pi_\mathrm{DH}$ and $\hat\Pi_\mathrm{DH}$, given by 
Eqs.~(\ref{eqn:pidh1}) and (\ref{eqn:pidhhat}), with 
the exact nonlinear version $\Pi$, given by Eq.~(\ref{eqn:pbmidplane}).
The dotted lines in Figures~\ref{figure:1}~and~\ref{figure:2}  
\textit{suggest} 
a better agreement between the linearized osmotic-pressure
difference $\hat\Pi_\mathrm{DH}$ and the full
nonlinear counterpart $\Pi$ 
 --- in comparison to the 
linearized version $\Pi_\mathrm{DH}$, obtained in Section~\ref{section:3}.
However, as shown below by \textit{explicit analytical} 
comparison, these \textit{numerical} evidences are in fact misleading. 

Asymptotic analytical expansions of the linearized osmotic-pressure difference 
$\hat\Pi_\mathrm{DH}$ about the weak-coupling $(\lb\to 0)$,
\bea
\hat\Pi_\mathrm{DH}&=&
\Pi_\mathrm{DH} +
k^2 \left[
  \frac{\eta^4}{16200}\left(\eta^2+5\right)\left(1-\eta^2\right) k^8 l^8
+{\mathrsfs O}\left({k^{10} l^{10}}\right) \right] =
k^2 \left[1-\frac{\eta^2}6{k^2 l^2} 
-\frac{\eta^2}{90}{\left(\eta^2-3\right) k^4 l^4} 
\right.\nonumber \qquad\\&&\left.
+\frac{\eta^2}{945}{\left(2\eta^2-5\right) k^6 l^6}
-\frac{\eta^2}{113400}{\left(7\eta^6+28\eta^4+\eta^2-84\right) k^8 l^8}
+{\mathrsfs O}\left({k^{10} l^{10}}\right)\right]-1,\qquad 
\eea
and the counterionic ideal-gas ($L\to 0$, finite $\Lambda$) limits, 
\bea
\hat\Pi_\mathrm{DH}&=& 
\Pi_\mathrm{DH} +
\frac{2}{\lambda l} \left\{\frac{\lambda^4}{675}
\left(\frac{l}{\lambda}\right)^6
+{\mathrsfs O}\left[ \left(l/\lambda\right)^7\right]\right\}, 
\eea
show explicitly that both linearized osmotic pressures, 
${\Pi}_\mathrm{DH}$ and
$\hat{\Pi}_\mathrm{DH}$, 
agree with the full nonlinear PB version $\Pi$ up to 
the \textit{same order} --- 
 cf.~Eqs.~(\ref{eqn:pblb})~to~(\ref{eqn:dhsmallsep}). Therefore, the 
\textit{numerical} indications of a better agreement of 
$\hat\Pi_\mathrm{DH}$, as suggested by Figures~\ref{figure:1} and
\ref{figure:2}, are purely \textit{fortuitous.}
In fact, for ratios $\Lambda/L> 10^2$ (beyond the values 
shown in Figure~\ref{figure:2}) one observes a
crossover between the deviations of the linearized 
versions, ${\Pi}_\mathrm{DH}$ and
$\hat{\Pi}_\mathrm{DH}$, with respect to the full nonlinear 
osmotic-pressure difference $\Pi$.

\begin{thebibliography}{99}
\bibitem{footnote1}
The Boltzmann-weighted equilibrium profiles~(\ref{eqn:planeprofile}) 
describe the ionic densities between the two charged plates. 
They do not apply to the infinite reservoir, that is 
only characterized by their associated 
(mean-field) ideal-gas microion chemical 
potentials,  $\beta\mu_\pm=\ln \left(\nb \zeta^3_\pm\right)$. 
The equilibrium density profiles~(\ref{eqn:planeprofile}) were
constructed in such a way to obey the charge-neutrality 
constraint~(\ref{eqn:neut_planes}), 
which implies an excess of counterions in the interplane region
to neutralize the negative charge of the plates. This counterion excess is
absent in the infinite salt reservoir.
See also Section 2 of Ref.~[\citen{sphere}] for an 
explanation of the distinct roles attributed to 
the Lagrange multiplier $\mu_\mathrm{el}$ 
and the microion chemical potentials $\mu_\pm$, which
are associated to different physical constraints of the problem,
namely, the overall charge-neutrality 
constraint~(\ref{eqn:neut_planes}) 
and the electrochemical equilibrium with the 
infinite salt reservoir of bulk density $\nb$.
\bibitem{footnote2}
Because of the frequent use of the standard gauge $(\mu_\mathrm{el}\equiv
0)$ the fact that this system constitutes 
a \textit{Donnan equilibrium problem} has not been emphasized.
\bibitem{footnote4}
Although the bare Coulomb interaction is 
long-ranged, even deionized aqueous suspensions at neutral pH  
characterized by small ionic strengths, $\nb\approx 10^{-7}\mbox{M}$, 
leads to Debye screening lengths of order $\kb^{-1}\approx 1\mu\mbox{m}\ll a$. 
Therefore, for surfaces with curvatures of macroscopic size  
the applicability of the Derjaguin approximation is
still valid.
\bibitem{footnote7}
We are presenting 
the expansions explicitly because algebraic-manipulation computer 
programs, like Mathematica and Maple, unfortunately do not perform 
series of the elliptic integrals with respect to the second argument
$m$.
\end{thebibliography}
\end{document}